\documentclass[12pt,onecolumn,twoside,draftclsnofoot,journal]{IEEEtran}
\usepackage{amsmath}
\usepackage{amssymb}
\usepackage{abbrevs}
\usepackage{subfig}
\usepackage{pstricks,pst-plot,pst-node}
\usepackage{hyperref}
\usepackage{units}
\usepackage{graphicx}
\usepackage[nomarkers,nofiglist,notablist]{endfloat}

\newtheorem{thm}{Theorem}
\newtheorem{lem}[thm]{Lemma}

\newcommand{\forloop}[5][1]%
{%
\setcounter{#2}{#3}%
\ifthenelse{#4}%
	{%
	#5%
	\addtocounter{#2}{#1}%
	\forloop[#1]{#2}{\value{#2}}{#4}{#5}%
	}%
	{%
	}%
}%

\newcommand{\TRext}{\ensuremath{TRext}}

\newcommand{\transpose}{\intercal}

\DeclareMathOperator{\sinc}{sinc}

\DeclareMathOperator{\rect}{rect}

\DeclareMathOperator*{\argmax}{arg\,max}
\DeclareMathOperator*{\argmin}{arg\,min}

\newcommand{\figureref}[1]{Fig.~\ref{#1}}
\newcommand{\figref}[1]{\figureref{#1}}
\newcommand{\tableref}[1]{Tab.~\ref{#1}}
\newcommand{\tabref}[1]{\tableref{#1}}

\newcommand{\sectionref}[1]{Section~\ref{#1}}

\newcommand{\secref}[1]{\sectionref{#1}}
\newcommand{\eqnref}[1]{Eqn.~(\ref{#1})}
\newcommand{\lemmaref}[1]{Lemma~\ref{#1}}

\newcommand\qh{\ensuremath{\hat{q}}}

\newcommand\vSh{\ensuremath{\mathbf{\hat{S}}}}

\newcommand\ve{\ensuremath{\mathbf{e}}}

\newcommand\vs{\ensuremath{\mathbf{s}}}

\newcommand\vH{\ensuremath{\mathbf{H}}}

\newcommand\vS{\ensuremath{\mathbf{S}}}

\newcommand\vV{\ensuremath{\mathbf{V}}}

\newcommand\vZ{\ensuremath{\mathbf{Z}}}

\newcommand\stilde{\ensuremath{\tilde{s}}}

\newabbrev\ASK{amplitude shift keying (ASK)}[ASK]
\newabbrev\AWGN{additive white Gaussian noise (AWGN)}[AWGN]
\newabbrev\BEP{bit error probability (BEP)}[BEP]
\newabbrev\BT{Binary Tree (BT)}[BT]
\newabbrev\BLF{backscatter link frequency (BLF)}[BLF]
\newabbrev\BCR{Batch Conflict Resolution (BCR)}[BCR]
\newabbrev\CRC{Cyclic Redundancy Check (CRC)}[CRC]
\newabbrev\CWT{Continous Wavelet Transform (CWT)}[CWT]
\newabbrev\DSB{Double Sideband (DSB)}[DSB]
\newabbrev\DSBASK{Double Sideband amplitude shift keying (DSB-ASK)}[DSB-ASK]
\newabbrev\EPC{electronic product code (EPC)}[EPC]
\newabbrev\EPCglobal{EPCglobal UHF Class--1 Generation--2 (EPC Gen2)}[EPC Gen2]
\newabbrev\FFT{fast Fourier transform (FFT)}[FFT]
\newabbrev\ISI{inter--symbol interference (ISI)}[ISI]
\newabbrev\LTI{linear time--invariant (LTI)}[LTI]
\newabbrev\ML{maximum likelihood (ML)}[ML]
\newabbrev\MAC{Media Access Control (MAC)}[MAC]
\newabbrev\MAP{maximum a--posteriori probability (MAP)}[MAP]
\newabbrev\MBT{Modified Binary Tree (MBT)}[MBT]
\newabbrev\MP{Matching Pursuit (MP)}[MP]
\newabbrev\MLSD{maximum likelihood sequence detector (MLSD)}[MLSD]
\newabbrev\PWM{Pulse Width Modulation (PWM)}[PWM]
\newabbrev\PW{Pulse Width (PW)}[PW]
\newabbrev\PC{Protocol Control (PC)}[PC]
\newabbrev\PIE{Pulse Interval Encoding (PIE)}[PIE]
\newabbrev\PRASK{Phase--Reversal Amplitude Shift Keying (PR-ASK)}[PR-ASK]
\newabbrev\RaCS{radar cross sections (RCS)}[RCS]
\newabbrev\RFID{Radio Frequency IDentification (RFID)}[RFID]
\newabbrev\RN{16--bit Random Number (RN16)}[RN16]
\newabbrev\SIC{Successive Interference Cancellation (SIC)}[SIC]
\newabbrev\SNR{signal--to--noise ratio (SNR)}[SNR]
\newabbrev\SSB{Single Sideband (SSB)}[SSB]
\newabbrev\SSBASK{Single Sideband amplitude shift keying (SSB-ASK)}[SSB-ASK]
\newabbrev\XPC{Extended Protocol Control (XPC)}[XPC]
\begin{document}

\title{Fast and Power Efficient Sensor Arbitration: Physical Layer Collision Recovery of Passive RFID Tags}

\author{Karsten Fyhn,~\IEEEmembership{Member,~IEEE,}
        Rasmus M. Jacobsen,~\IEEEmembership{Member,~IEEE,}
        Petar Popovski,~\IEEEmembership{Senior Member,~IEEE,}
        Anna Scaglione,~\IEEEmembership{Fellow,~IEEE,}
        and~Torben Larsen,~\IEEEmembership{Senior Member,~IEEE}%
        \thanks{The authors e--mails are: \{kfyhn,raller,petarp,tl\}@es.aau.dk, and ascaglione@ucdavis.edu. 
        The first two authors are students at the Aalborg University elite--programme, \url{http://eliteeducation.aau.dk}.}%
}

\maketitle

\begin{abstract}
% Introduction
This work concerns physical layer collision recovery for cheap sensors with allowed variations in frequency and delay of their communications.
The work is presented as a generic, communication theoretic framework and demonstrated using UHF RFID tag technology.
Previous work in this area has not provided recovery for more than two tags,
which is shown to be possible in this work.
Also presented is a novel mathematical model of the tag signal,
incorporating the allowed variations in frequency and delay. 

% Method
The main motivation is seen in the observation that random variations in frequency and delay make the collided signals of different tags separable.
The collision recovery is done by estimating the sensor specific variation in frequency and delay and
using these estimates in a successive interference cancellation algorithm and a maximum likelihood sequence decoder,
to iteratively reconstruct a sensor signal and remove it from the received signal.

% Results
Numerical simulations show that the estimates and proposed algorithm are effective in recovering collisions.
The proposed algorithm is then incorporated into a numerical simulation of the $Q$--protocol for UHF RFID tags and is shown to be effective in providing fast and power efficient sensor arbitration.

% Analysis
% Discussion
\end{abstract}


%%%%%%%%%%%%%%%%%%%%%%
\section{Introduction}
%%%%%%%%%%%%%%%%%%%%%%
%\begin{itemize}
%  \item The RFID communication paradigm - tags collide, Q protocol - a signal space representation is defined
%  \item Multiple tag decoding would be beneficial, implicit cardinality estimation $=$ improvement to the Q protocol 
%  \item Current state--of--the--art in multiple tag decoding and cardinality estimation
%  \item Why joint decoding is impossible - frequency and delay variation - relate to wavelet theory!
%  \item Variation in frequency and delay may be used to differentiate and decode multiple tags
%  \item If multiple tag decoding is incorporated into the Q-protocol, a tag set may be resolved faster and with fewer slots
%\end{itemize}
In the current standard for UHF \RFID,  
the protocol imposes a simple tag--to--reader communication to allow for a simple tag structure \cite{epc}.
Collisions occur at the reader when multiple tags simultaneously reply to a query sent from a reader.
To combat this, a range of anti--collision, or arbitration protocols have been designed, which ensure that all tags at some point during arbitration are queried individually.

\begin{figure}
\begin{center}
\includegraphics[scale=0.9]{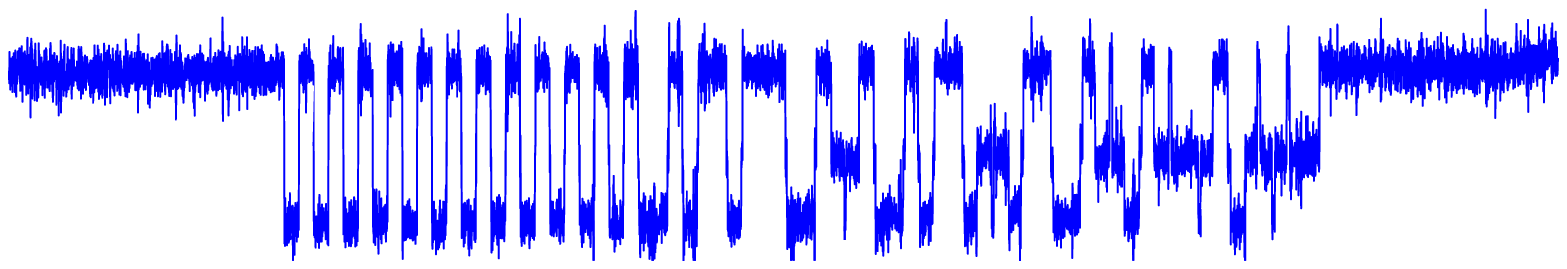}
%\psgrid(0,-1)(-15,3)
\rput(-11.92,-0.4){%
  \psline[linecolor=red,linestyle=dashed](0,0)(0,0.4)
  \psline[linecolor=red,linestyle=dashed](-2.5,0)(-2.5,0.4)
  \psline[linecolor=red,linestyle=dashed](-2.5,0.2)(0,0.2)
  \uput[180](-0.55,-0.15){\small\textcolor{red}{Delay}}
}
\rput(-10.565,-0.10){%
  \psline[linecolor=red,linestyle=dashed](0,0)(0,2.3)
  \psline[linecolor=red,linestyle=dashed](0.27,0)(0.27,2.3)
  \rput(0.1,2.7){\small\textcolor{red}{$\frac{1}{\text{Link frequency}}$}}
}
\pscircle[linecolor=red](-4.35,2.0){0.2}
\pscircle[linecolor=red](-3.82,1.9){0.2}
\pscircle[linecolor=red](-2.85,2.0){0.3}
\end{center}
\caption{Signal level of a collided tag signal, with two participating tags, as measured by a reader. The nominal link frequency (BLF) is $44.44\unit{kHz}$, which is the reason for the small delay difference.
	The tags are synchronized to begin with, but differs later in the communication, as shown by the red circles.}
\label{fig:delay_and_link_freq}
\end{figure}

Since tag production is targeting low prices, large variations are allowed on the parameters employed by a tag to modulate the backscattered signal. 
With the existing decoding techniques, the reader uses a coherent structure to \emph{mitigate the problem} of varying tag encoding parameters \cite{adv_sync_and_decod}. 
On the other hand, in this work, we analyze a new communication theoretic framework, 
in which such a variation across tags is considered an \emph{enabler} allowing to differentiate tag information in a collided tag reply.
This may be seen as \emph{cheap CDMA}, where the code separation between tags is generated due to production tolerances.
The two parameters that vary across tags are the link frequency and the time of reply, see \figref{fig:delay_and_link_freq}.

Link frequency exhibits the most significant variation. That is the frequency used by a tag to encode a reply during backscatter onto the amplitude modulated carrier from the reader. The tolerance limits for this frequency, defined in \EPCglobal \cite{epc}, allow up to $\pm$22\% variation per message from the \emph{nominal} link frequency, denoted the \BLF.
Also, to allow for minimal synchronization at the tag side, the time where a tag initiates a reply is also allowed to vary.
The range can be as large as $24\unit{\mu s}$ for some \BLF{}s which is equivalent to the duration of several encoded symbols.

In this work, we show how these two parameters may be estimated and used for decoding multiple tag replies in a single slot.
To the best of our knowledge, this is the first method to achieve decoding of more than two UHF RFID tags by using diversification of the tag parameters.
In \cite{sep_of_multi_tags} the authors show how to decode up to four LF tags using signal constellations in the I/Q domain.
Their method assumes centralized, reader--controlled link frequency, which is not valid for UHF tags 
due to the allowed frequency variation.
For UHF tags it is shown in \cite{rfid_reader_recv} how up to two tags can be decoded,
also using the signal constellations in the I/Q domain.
This is done using zero forcing and successive interference cancellation.
Neither work uses \MLSD to improve the decoding.
Other work uses multiple antennas for separation of multiple tags \cite{smart_antennas,smart_antennas2,antenna_array},
where our work is based on single antenna systems.
Also, a related field is tag population estimation such as in \cite{ML_tag_pop_est,extract_info_from_tag_col,conflict_multiplicity}.
Similar methods of code separation are also used in these works for estimating the number of tags in a collision.

One contribution of this work is a detailed mathematical signal and channel model of tag to reader communication in UHF RFID communication (described in \secref{sec:sig_chan_model}) and
an estimator of link frequency and delay of a tag reply (derived in \secref{sec:estimator}).
This estimator is then shown to be useful for multiple tag decoding, when used together with the $Q$--protocol from \EPCglobal (shown in Sections \ref{sec:data_decoding}, \ref{sec:q_protocol} and \ref{sec:results}).
This protocol is used for arbitration of tags and is an ALOHA--based protocol, which resolves tags in slots.
A slot can be either Single, Collision or Idle. Only when a slot is Single is a tag resolved.
Using our work, we show the gain of being able to resolve some of the Collision slots as well.
In short, our communication theoretic analysis of the RFID channel allows us to transform the tag synchronization diversity from a foe to a friend,
and show that in the UHF RFID case, the variations in the sensor encoding parameters may be used with a \MLSD to enhance the decoding of multiple tags.
We show that a significant gain is achievable when using our method in the $Q$--protocol. Note that the ideas presented here are applicable beyond UHF RFID, to a wide range of scenarios with cheap, passive, clock-less tags and sensors. 

\begin{figure}
\centering
\includegraphics[width=0.8\textwidth]{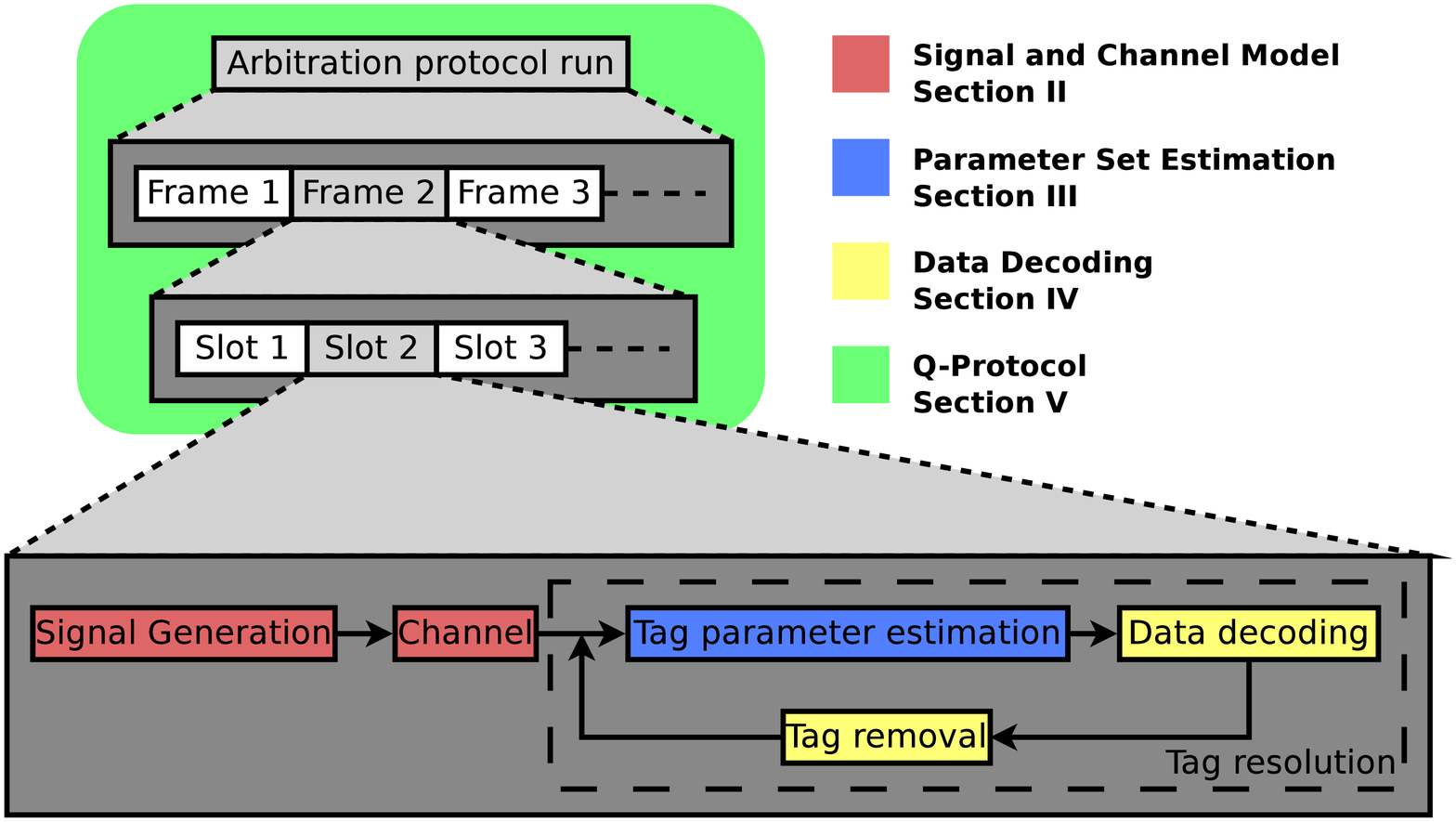}
\caption{An overview of the following sections. The iterative process of tag resolution is also described together with data decoding in \secref{sec:data_decoding}.}
\label{fig:overview}
\end{figure}

The structure of thepaper is visualized in \figref{fig:overview}.
The tag resolution process is iterative because the parameter estimation process 
obtains the parameters of one tag at a time -- the strongest in the collided signal.
The process therefore consists of estimating the parameters of the strongest tag, decoding its data, 
subtracting its signal contribution and reiterating the process to find the next strongest tag.
This continues until there are no more tags present in the residual signal.

%The following four sections are as shown in \figref{fig:overview} and are described bottom-to-top.
%The received, possibly collided signal at the reader is modelled in \secref{sec:sig_chan_model},
%where the channel and the basis functions for FM0 and Miller encoding schemes specify the received signal.
%This is followed, in \secref{sec:estimator},
%by the estimation framework for obtaining the link frequency and delay for a tag in the reply.
%
%Data decoding follows in \secref{sec:data_decoding} for multiple tag decoding.
%Clearly, tag cardinality estimation may be treated as a bi--product and an implicit part of multiple tag decoding.
%
%The reason why the process of tag resolution is iterative, is that the parameter estimation process 
%obtains the parameters of one tag at a time -- the strongest in the collided signal.
%The process therefore consists of estimating the parameters of the strongest tag, decoding its data, 
%subtracting its signal contribution and reiterate the process to find the next most strongest tag.
%This process continues until no more tags are present in the residual signal.
%
%In \secref{sec:q_protocol}, we show how the $Q$--protocol has been implemented and which assumptions has been taken,
%followed by the results of our numerical simulations in \secref{sec:results}.
%Finally, in \secref{sec:conclusion}, the paper is concluded.

%%%%%%%%%%%%%%%%%%%%%%%%%%%%%%%%%%
\section{Signal and Channel Model}
\label{sec:sig_chan_model}
%%%%%%%%%%%%%%%%%%%%%%%%%%%%%%%%%%
%\begin{itemize}
%  \item Basis functions and memory matrices
%  \item Tag control signal definition
%  \item Channel model
%\end{itemize}
This section describes the mathematical framework we have derived for representing tag signals
and the channel model we employ to simulate their transmission over the air.
We begin with the derivation of basis functions and the signal space representation of UHF RFID tag to reader communication,
which is based on either FM0 or Miller encoding \cite{epc}.
As tag to reader communication is based on backscattering \cite{finkenzeller} of a carrier wave,
the tag transmission signal should be seen as a \emph{control signal},
specifying whether a tag backscatters the carrier wave or not.

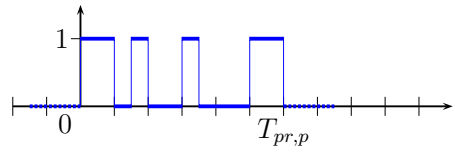
\begin{figure}[h]
\centering
\psscalebox{0.9}{%
\begin{pspicture}(-2,-0.5)(5.5,1.5)
\psaxes[dx=0.5,labels=none]{->}(0,0)(-1,0)(5.5,1.5)
\uput[180](0,1){$1$}
\uput[-135](0,0){$0$}
\uput[-90](3,0){$T_{pr,p}$}
\psscalebox{0.25 1}{%
\psline[linewidth=2\pslinewidth,linecolor=blue](0,0)(0,1)(2,1)(2,0)(3,0)(3,1)(4,1)(4,0)(6,0)(6,1)(7,1)(7,0)(10,0)(10,1)(12,1)(12,0)
\psline[linewidth=2\pslinewidth,linecolor=blue,linestyle=dashed](-3,0)(0,0)
\psline[linewidth=2\pslinewidth,linecolor=blue,linestyle=dashed](12,0)(15,0)
}
\end{pspicture}}%
\caption{FM0 preamble control signal with $\TRext=0$ (short preamble). The bit sequence is $\{1,0,1,0,v,1\}$, 
where the $v$ is a symbol breaking the encoding (more on this later). 
The ticks on the x-axis denote a symbol duration.}
\label{fig:tag_signal_before_channel}
\end{figure}

An example of the control signal for the short preamble in FM0 encoding is shown in \figref{fig:tag_signal_before_channel}.
This example is used in the remainder of this section to explain the signal encoding.
We first define the basis functions and signal waveforms used to generate each individual symbol. Then we describe
the state machine that generates sequences of FM0 and Miller symbols.

%%%%%%%%%%%%%%%%%%%%%%%%%%%%%%%%%%%%%%%%%%%%%%%%%%%%%%%%%
\subsection{Basis Functions for FM0 and Miller Encoding}
\label{sec:basis_functions}
%%%%%%%%%%%%%%%%%%%%%%%%%%%%%%%%%%%%%%%%%%%%%%%%%%%%%%%%%

Let $\mathcal{M}_p=\{m_0,m_1,\ldots,m_{N_{\mathcal{M}}-1}\}, m_n\in\{0,1\}$ be the data message backscattered by tag $p$ after a Query, QueryAdjust, or QueryRep command, not including pre-- and postamble.
This message is the reply message in a slot during arbitration with the $Q$--protocol.
It contains a 16--bit random number (RN16), and so $N_{\mathcal{M}}=16$.
To transmit this message, a tag first encodes it using FM0 or Miller after which it is backscattered to the reader,
through the channel.
The FM0 and Miller basis functions are not rigorously defined in \cite{epc},
but the signal waveforms for the respective encoding schemes are specified.

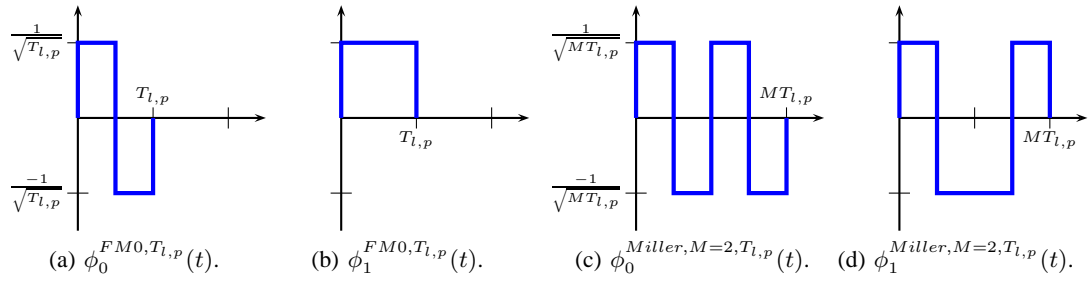
\begin{figure}[h!]
\centering
\subfloat[$\phi_0^{FM0,T_{l,p}}(t)$.]{%
\begin{pspicture}(-1,-1.5)(2.5,1.5)
\tiny
\psaxes[labels=none]{->}(0,0)(0,-1.5)(2.5,1.5)
\uput[180](0,1){$\frac{1}{\sqrt{T_{l,p}}}$}
\uput[180](0,-1){$\frac{-1}{\sqrt{T_{l,p}}}$}
\uput[90](1,0){$T_{l,p}$}
\psline[linewidth=2\pslinewidth,linecolor=blue](0,0)(0,1)(0.5,1)(0.5,-1)(1,-1)(1,0)
\end{pspicture}%
}
\subfloat[$\phi_1^{FM0,T_{l,p}}(t)$.]{%
\begin{pspicture}(-1,-1.5)(2.5,1.5)
\tiny
\psaxes[labels=none]{->}(0,0)(0,-1.5)(2.5,1.5)
\uput[-90](1,0){$T_{l,p}$}
\psline[linewidth=2\pslinewidth,linecolor=blue](0,0)(0,1)(1,1)(1,0)
\end{pspicture}%
}
\quad
\subfloat[$\phi_0^{Miller,M=2,T_{l,p}}(t)$.]{%
\begin{pspicture}(-1,-1.5)(2.5,1.5)
\tiny
\psaxes[dx=0.5,labels=none]{->}(0,0)(0,-1.5)(2.5,1.5)
\uput[180](0,1){$\frac{1}{\sqrt{MT_{l,p}}}$}
\uput[180](0,-1){$\frac{-1}{\sqrt{MT_{l,p}}}$}
\uput[90](2,0){$MT_{l,p}$}
\psline[linewidth=2\pslinewidth,linecolor=blue](0,0)(0,1)(0.5,1)(0.5,-1)(1,-1)(1,1)(1.5,1)(1.5,-1)(2,-1)(2,0)
\end{pspicture}%
}
\subfloat[$\phi_1^{Miller,M=2,T_{l,p}}(t)$.]{%
\begin{pspicture}(-1,-1.5)(2.5,1.5)
\tiny
\psaxes[dx=0.5,labels=none]{->}(0,0)(0,-1.5)(2.5,1.5)
\uput[-90](2,0){$MT_{l,p}$}
\psline[linewidth=2\pslinewidth,linecolor=blue](0,0)(0,1)(0.5,1)(0.5,-1)(1.5,-1)(1.5,1)(2,1)(2,0)
\end{pspicture}%
}
\caption{Basis functions for FM0 ($M=1$) and Miller with $M=2$.}
\label{fig:fm0_miller_basis_functions}
\end{figure}

Let $\phi_k^T(t)$, $k=0,1$ be basis functions having \emph{support duration} $MT$.
That is $\phi_k^T(t)=0$ for $t<0$ and $t>MT$,
where $M$ is a symbol period multiplier.
For FM0, $M=1$, and the basis functions are:
\begin{equation}
\nonumber\phi_0^{FM0,T}(t) =\frac{1}{\sqrt{T}}\left\{\rect\left(\frac{t-\frac{T}{4}}{\frac{T}{2}}\right)-\rect\left(\frac{t-\frac{3T}{4}}{\frac{T}{2}}\right)\right\} \qquad
\label{eqn:phiFM0}\phi_1^{FM0,T}(t) =\frac{1}{\sqrt{T}}\rect\left(\frac{t-\frac{T}{2}}{T}\right),
\end{equation}
where the uses of $\rect(\cdot)$ are scaled so the bases have unit energy\footnote{%
Notice that the basis functions do \emph{not} have zero mean, 
i.e. they correlate highly with the readers carrier wave echo.
However, because the tag signal has overall zero mean, as is shown later,
this is not a problem.
}.
The tag signal is generated here with ideal on--off keying.
Noise and the hardware limitations do not allow such an instantaneous transition,
however the difference is considered negligible, as in e. g. \cite{sep_of_multi_tags,single_ant_phy_collision_recov}.
For Miller, $M=2,4,8$, corresponding to the number of subcarrier cycles in the basis function:
\begin{align}
\nonumber\phi_0^{Miller,M,T}(t)&=\frac{1}{\sqrt{MT}}
\sum_{j=0}^{M-1}\left[\rect\left(\frac{t-\left(j+\frac{1}{4}\right)T}{\frac{T}{2}}\right)-\rect\left(\frac{t-\left(j+\frac{3}{4}\right)T}{\frac{T}{2}}\right)\right],\\
\nonumber\phi_1^{Miller,M,T}(t)&=\frac{1}{\sqrt{MT}}\Bigg\{
\sum_{j=0}^{\frac{M}{2}-1}\left[\rect\left(\frac{t-\left(j+\frac{1}{4}\right)T}{\frac{T}{2}}\right)-\rect\left(\frac{t-\left(j+\frac{3}{4}\right)T}{\frac{T}{2}}\right)\right]\\
&\quad\label{eqn:phiMiller}\quad-\sum_{j=\frac{M}{2}}^{M-1}\left[\rect\left(\frac{t-\left(j+\frac{1}{4}\right)T}{\frac{T}{2}}\right)-\rect\left(\frac{t-\left(j+\frac{3}{4}\right)T}{\frac{T}{2}}\right)\right]\Bigg\}.
\end{align}
The basis functions are shown in \figureref{fig:fm0_miller_basis_functions},
when evaluated for the \emph{tag dependent subcarrier period} defined from the tag dependent link frequency as $T_{l,p}=\frac{1}{f_{l,p}}$,
where $f_{l,p}$ is the link frequency for tag $p$ where the allowed frequency variation is included.

%%%%%%%%%%%%%%%%%%%%%%%%%%%%%%%%%%%%%%%%%%%%%%%%%%%%%%%%%
\subsection{Possible Signal Waweforms using Basis Functions}
\label{sec:signal_waveforms}
%%%%%%%%%%%%%%%%%%%%%%%%%%%%%%%%%%%%%%%%%%%%%%%%%%%%%%%%%

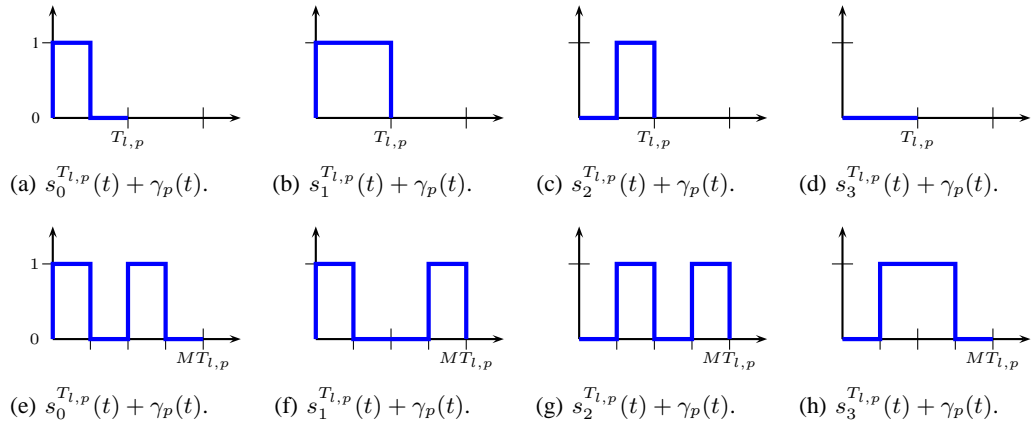
\begin{figure}[h!]
\centering
\subfloat[$s_0^{T_{l,p}}(t)+\gamma_p(t)$.]{%
\begin{pspicture}(-1,-0.5)(2.5,1.5)
\tiny
\psaxes[labels=none]{->}(0,0)(2.5,1.5)
\uput[180](0,1){1}
\uput[180](0,0){0}
\uput[-90](1,0){$T_{l,p}$}
\psline[linewidth=2\pslinewidth,linecolor=blue](0,0)(0,1)(0.5,1)(0.5,0)(1,0)
\end{pspicture}%
}
\subfloat[$s_1^{T_{l,p}}(t)+\gamma_p(t)$.]{%
\begin{pspicture}(-1,-0.5)(2.5,1.5)
\tiny
\psaxes[labels=none]{->}(0,0)(2.5,1.5)
\uput[-90](1,0){$T_{l,p}$}
\psline[linewidth=2\pslinewidth,linecolor=blue](0,0)(0,1)(1,1)(1,0)
\end{pspicture}%
}
\subfloat[$s_2^{T_{l,p}}(t)+\gamma_p(t)$.]{%
\begin{pspicture}(-1,-0.5)(2.5,1.5)
\tiny
\psaxes[labels=none]{->}(0,0)(2.5,1.5)
\uput[-90](1,0){$T_{l,p}$}
\psline[linewidth=2\pslinewidth,linecolor=blue](0,0)(0,0)(0.5,0)(0.5,1)(1,1)(1,0)
\end{pspicture}%
}
\subfloat[$s_3^{T_{l,p}}(t)+\gamma_p(t)$.]{%
\begin{pspicture}(-1,-0.5)(2.5,1.5)
\tiny
\psaxes[labels=none]{->}(0,0)(2.5,1.5)
\uput[-90](1,0){$T_{l,p}$}
\psline[linewidth=2\pslinewidth,linecolor=blue](0,0)(1,0)
\end{pspicture}%
}\\
\subfloat[$s_0^{T_{l,p}}(t)+\gamma_p(t)$.]{%
\begin{pspicture}(-1,-0.5)(2.5,1.5)
\tiny
\psaxes[dx=0.5,labels=none]{->}(0,0)(2.5,1.5)
\uput[180](0,1){$1$}
\uput[180](0,0){$0$}
\uput[-90](2,0){$MT_{l,p}$}
\psline[linewidth=2\pslinewidth,linecolor=blue](0,0)(0,1)(0.5,1)(0.5,0)(1,0)(1,1)(1.5,1)(1.5,0)(2,0)
\end{pspicture}%
}
\subfloat[$s_1^{T_{l,p}}(t)+\gamma_p(t)$.]{%
\begin{pspicture}(-1,-0.5)(2.5,1.5)
\tiny
\psaxes[dx=0.5,labels=none]{->}(0,0)(2.5,1.5)
\uput[-90](2,0){$MT_{l,p}$}
\psline[linewidth=2\pslinewidth,linecolor=blue](0,0)(0,1)(0.5,1)(0.5,0)(1.5,0)(1.5,1)(2,1)(2,0)
\end{pspicture}%
}
\subfloat[$s_2^{T_{l,p}}(t)+\gamma_p(t)$.]{%
\begin{pspicture}(-1,-0.5)(2.5,1.5)
\tiny
\psaxes[dx=0.5,labels=none]{->}(0,0)(2.5,1.5)
\uput[-90](2,0){$MT_{l,p}$}
\psline[linewidth=2\pslinewidth,linecolor=blue](0,0)(0.5,0)(0.5,1)(1,1)(1,0)(1.5,0)(1.5,1)(2,1)(2,0)
\end{pspicture}%
}
\subfloat[$s_3^{T_{l,p}}(t)+\gamma_p(t)$.]{%
\begin{pspicture}(-1,-0.5)(2.5,1.5)
\tiny
\psaxes[dx=0.5,labels=none]{->}(0,0)(2.5,1.5)
\uput[-90](2,0){$MT_{l,p}$}
\psline[linewidth=2\pslinewidth,linecolor=blue](0,0)(0.5,0)(0.5,1)(1.5,1)(1.5,0)(2,0)
\end{pspicture}%
}
\caption{Control waveforms for FM0 (top) and Miller with $M=2$ (bottom) where $s_0(t)$ and $s_2(t)$ encodes symbol--0 and $s_1(t)$ and $s_3(t)$ symbol--1.
$\gamma_p(t)=\frac{1}{2}\rect\left(\frac{t-\frac{MT_{l,p}}{2}}{MT_{l,p}}\right)$ is the offset added in the support duration.}
\label{fig:control_signal_waveforms}
\end{figure}

RFID tags modulate the received waveform by alternating the control signal between two states: 0 (OFF) and 1 (ON).
OFF means that the tag absorbs the power it receives and ON means it reflects it back.
The control signal is generated in two steps:
\begin{enumerate}
	\item Signal waveforms are derived from the encoding dependent basis functions in Eqns. \ref{eqn:phiFM0} and \ref{eqn:phiMiller} with signal levels $\pm\frac{1}{2}$.
	\item A constant offset of $\frac{1}{2}$ is added to the encoded message to create the control signal.
\end{enumerate}
In the first step, a set of signal waveforms are found as a linear combination of the basis functions.
This is accomplished using the following signal space representation:
\begin{align}
  \vV_{\mathcal{E}}=\frac{\sqrt{MT_{l,p}}}{2}\vV,\qquad\vV=
\begin{bmatrix}
1&0&-1&0\\
0&1&0&-1
\end{bmatrix},
\label{eqn:sig_space_rep}
\end{align}
This matrix we shall use in later sections as a dictionary for possible signals.
Using this signal space representation, the possible signal waveforms are generated as:
\begin{align}
s_i^{T_{l,p}}(t)&=\sum_{j=0}^{1} v_{j,i}\phi_j^{T_{l,p}}(t),
\end{align}
where $v_{j,i}$ picks out an element from $\vV_{\mathcal{E}}$ in \eqnref{eqn:sig_space_rep}, 
$\phi_j^{T_{l,p}}(t)$, $j=0,1$ are the encoding dependent basis functions and where the signal level is $-\frac{1}{2}$ when the tag absorbs and $\frac{1}{2}$ when it reflects.
Adding the signal level offset $\frac{1}{2}$ to these signal waveforms in their support duration gives the control waveforms illustrated in \figureref{fig:control_signal_waveforms}.

The control waveforms now allow us to generate single symbols.
The following describes how symbol sequences are generated using the memory in the encoding schemes.
We exploit this memory later in the decoding, which significantly improves the decoding.

%%%%%%%%%%%%%%%%%%%%%%%%%%%%%%%%%%%%%%%%%%%%%%%%%%%%%%%%%%%%%%%%%%%%%%%%%%%%
\subsection{Generating the Control Signal using the Inherent Encoding Memory}
%%%%%%%%%%%%%%%%%%%%%%%%%%%%%%%%%%%%%%%%%%%%%%%%%%%%%%%%%%%%%%%%%%%%%%%%%%%%
\label{sec:control_signal}

An important property for FM0 and Miller is the inherent memory,
where the signal waveforms used for encoding of the symbol $m_n$ depends on the previously sent symbol $m_{n-1}$.
Let the signal waveforms $s_k^{T_{l,p}}(t)$ correspond to \emph{state} $s_k$,
then the state machine for FM0 and Miller is in \figref{fig:fm0_miller_state_diagrams},
from which we obtain the symbol--dependent transition matrices $\vH_{m_n}$, $m_n=\{0,1\}$:

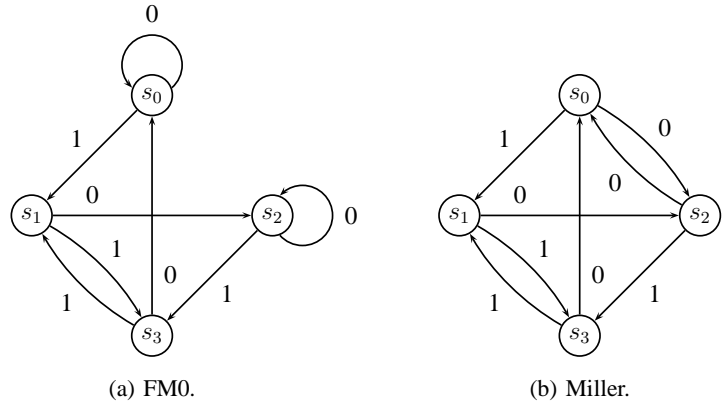
\begin{figure}[h1]
\centering
\subfloat[FM0.]{%
\scalebox{0.8}{%
\begin{pspicture}(-0.5,-0.5)(4.5,5.5)
\cnodeput(2,4){s0}{$s_0$}
\cnodeput(0,2){s1}{$s_1$}
\cnodeput(4,2){s2}{$s_2$}
\cnodeput(2,0){s3}{$s_3$}
\psset{arrows=->,arcangle=15}
\def\radius{0.5}
\nccircle{s0}{\radius}\nbput{0}
\ncline{s0}{s1}\nbput{1}
\ncline{s1}{s2}\naput[npos=0.2]{0}
\ncarc{s1}{s3}\naput{1}
\nccircle[angleA=-90]{s2}{\radius}\nbput{0}
\ncline{s2}{s3}\naput{1}
\ncline{s3}{s0}\nbput[npos=0.2]{0}
\ncarc{s3}{s1}\naput{1}
\end{pspicture}%
}}
\qquad\qquad
\subfloat[Miller.]{%
\scalebox{0.8}{%
\begin{pspicture}(-0.5,-0.5)(4.5,4.5)
\cnodeput(2,4){s0}{$s_0$}
\cnodeput(0,2){s1}{$s_1$}
\cnodeput(4,2){s2}{$s_2$}
\cnodeput(2,0){s3}{$s_3$}
\psset{arrows=->,arcangle=15}
\ncline{s0}{s1}\nbput{1}
\ncarc{s0}{s2}\naput{0}
\ncline{s1}{s2}\naput[npos=0.2]{0}
\ncarc{s1}{s3}\naput{1}
\ncarc{s2}{s0}\naput{0}
\ncline{s2}{s3}\naput{1}
\ncline{s3}{s0}\nbput[npos=0.2]{0}
\ncarc{s3}{s1}\naput{1}
\end{pspicture}%
}}
\caption{State diagrams for FM0 and Miller encoding. A 0 and 1 indicates the symbol sent for the transition to take place, and $s_k$ indicates the state representing the signal waveform $s_k^{T_{l,p}}(t)$ used to encode a symbol.}
\label{fig:fm0_miller_state_diagrams}
\end{figure}

\begin{align}
\vH_{m_n=0}^{FM0} &= 
\begin{bmatrix}
	1 & 0 & 0 & 1 \\
	0 & 0 & 0 & 0 \\
	0 & 1 & 1 & 0 \\
	0 & 0 & 0 & 0 
\end{bmatrix},
&\vH_{m_n=1}^{FM0} &=
\begin{bmatrix}
	0 & 0 & 0 & 0 \\
	1 & 0 & 0 & 1 \\
	0 & 0 & 0 & 0 \\ 
	0 & 1 & 1 & 0 
\end{bmatrix},\\
\vH_{m_n=0}^{Miller} &= 
\begin{bmatrix}
	0 & 0 & 1 & 1 \\
	0 & 0 & 0 & 0 \\
	1 & 1 & 0 & 0 \\
	0 & 0 & 0 & 0 
\end{bmatrix},
&\vH_{m_n=1}^{Miller} &=
\begin{bmatrix}
	0 & 0 & 0 & 0 \\
	1 & 0 & 0 & 1 \\
	0 & 0 & 0 & 0 \\
	0 & 1 & 1 & 0 
\end{bmatrix},
\label{eqn:H_matrices}
\end{align}
where the $(k,k')$th entry equal to $1$ indicates a valid transition from state $s_{k'-1}$ to state $s_{k-1}$.
Also, let $\vS_{\mathcal{M}_p}$ be a $4\times N_{\mathcal{M}}$ state select matrix generated using $\vH_0$, $\vH_1$ and the message $\mathcal{M}_p$.
Each column vector $\vs_{\mathcal{M}_p,n}$ in $\vS_{\mathcal{M}_p}$ is one of the coordinate column vectors $\ve_k$,
where $k=1,2,3,4$ denotes the state $s_0$, $s_1$, $s_2$, or $s_3$, respectively, used to encode the $n$th symbol in $\mathcal{M}_p$:
\begin{align}
\label{eqn:S_matrix}\vS_{\mathcal{M}_p} &= \begin{bmatrix} \vs_{\mathcal{M}_p,0} & \vs_{\mathcal{M}_p,1} & \cdots & \vs_{\mathcal{M}_p,{N_{\mathcal{M}}-1}} \end{bmatrix}
= \begin{bmatrix} \vH_{m_0}\vs_{\text{init}} & \vH_{m_1}\vs_0 & \cdots & \vH_{m_{N_{\mathcal{M}}-1}}\vs_{N_{\mathcal{M}}-2} \end{bmatrix},
\end{align}
where $\vs_{\text{init}}$ denotes the state prior to the first symbol in $\mathcal{M}_p$.
This state follows from the last symbol in the preamble.

An example is the state select matrix used to generate the signal in \figref{fig:tag_signal_before_channel}:
\begin{align*}
\vS_{\mathcal{M}_p} &=
\begin{bmatrix}
	1 & 0 & 0 & 0 & 0 & 1 \\
	0 & 0 & 0 & 1 & 0 & 0 \\
	0 & 1 & 0 & 0 & 0 & 0 \\
	0 &	0 &	1 &	0 &	1 &	0
\end{bmatrix}
\end{align*}
From \EPCglobal it is known that the state for the last transmitted symbol in the preamble is $s_1$ for FM0 and $s_3$ for Miller,
and the respective initialization vectors are:
\begin{align}
\vs_{\text{init}}^{FM0} = \ve_2,\qquad\text{and}\qquad \vs_{\text{init}}^{Miller} = \ve_4.
\label{eqn:s_init}
\end{align}
%\begin{figure}
%\centering
%\begin{pspicture}(5,5)
%%\psgrid
%\rput(2,2){%
%  \psaxes[labels=none,ticks=none]{<->}(0,0)(-1.5,-1.5)(1.5,1.5)
%  \psdots(1,0)(0,1)(-1,0)(0,-1)
%  \uput[-90](1,0){$\vv_{\mathcal{E},0}$}
%  \uput[0](0,1){$\vv_{\mathcal{E},1}$}
%  \uput[-90](-1,0){$\vv_{\mathcal{E},2}$}
%  \uput[0](0,-1){$\vv_{\mathcal{E},3}$}
%}
%\end{pspicture}
%\label{fig:sig_constellation}
%\caption{The signal space constellation.}
%\end{figure}
The control signal waveform describing the message part for tag $p$ directly follows as:
\begin{align}
\nonumber c_{\mathcal{M}_p}(t) &=\sum_{n=0}^{N_{\mathcal{M}}-1}\sum_{k=0}^3 \ve_{k+1}^\transpose\vS_{\mathcal{M}_p}\ve_{n+1}s_k^{T_{l,p}}(t-nMT_{l,p})+\gamma_p(t),\\
\label{eqn:signal_enc}&=\sum_{n=0}^{N_{\mathcal{M}}-1}\sum_{k=0}^1 \ve_{k+1}^\transpose\vS_{\mathcal{M}_p}\ve_{n+1}\vV_{\mathcal{E}}\phi_k^{T_{l,p}}(t-nMT_{l,p})+\gamma_p(t),
\end{align}
where $D_{\mathcal{M}_p}=T_{l,p}MN_{\mathcal{M}}$ is the duration of the data message,
and $\gamma_p(t)=\frac{1}{2}\rect\left(\frac{t-\frac{D_{\mathcal{M}_p}}{2}}{D_{\mathcal{M}_p}}\right)$
adds the offset ensuring that the control signal has signal levels 0 or 1.

Pre-- and postamble control signals are added to the message control signal,
where the preamble depends on whether FM0 or Miller is used for encoding.
Let $c_{pr,p}(t)$ be the preamble control signal generated with tag link frequency $f_{l,p}$, and let $D_{pr,p}$ be the support duration of the preamble.
Also, let $c_{po,p}(t)$ be the postamble control signal with support duration equivalent to the duration of one symbol and the complete transmitted signal from tag $p$ is:
\begin{subequations}
\label{eqn:c_p}
\begin{align}
c_p(t) &= c_{pr,p}(t) + c_{\mathcal{M},p}(t-D_{pr,p}) + c_{po,p}\left(t-D_{pr,p}-D_{\mathcal{M}_p}\right).
\label{eqn:c_p1}
\end{align}
In the following sections an alternative representation of \eqnref{eqn:c_p1} is used where the message and the postamble symbol are included in a combined structure.
It is known from \EPCglobal that the postamble symbol is symbol--1 which corresponds to state $s_1$ or $s_3$ depending on the state of the last encoded symbol in $\mathcal{M}_p$.
Let the state select matrix $\vS_p$ be $4\times(N_{\mathcal{M}}+1)$ where the last entry is for the postamble symbol:
\[ \vS_p = \begin{bmatrix}\vS_{\mathcal{M}_p}&\vH_1\vs_{N_{\mathcal{M}}-1}\end{bmatrix} \]
which follows from \eqnref{eqn:S_matrix} and \eqnref{eqn:c_p1} is rewritten from an extended version of \eqnref{eqn:signal_enc}:
\begin{align}
\nonumber c_p(t) &= c_{pr,p}(t) + 
\sum_{n=0}^{N_{\mathcal{M}}}\sum_{k=0}^3 \ve_{k+1}^\transpose\vS_p\ve_{n+1}s_k^{T_{l,p}}(t-D_{pr,p}-nMT_{l,p}) + \gamma_p(t),
\end{align}
\end{subequations}
where the support duration of $\gamma_p(t)$ is increased to include the postamble symbol $
\gamma_p(t)=\frac{1}{2}\rect\left(\frac{t-D_{pr,p}-\frac{T_{l,p}M\left(N_{\mathcal{M}}+1\right)}{2}}{T_{l,p}M\left(N_{\mathcal{M}}+1\right)}\right).
$

%%%%%%%%%%%%%%%%%%%%%%%%%%%%%%%%%%%%%%%%%%%%%%%%%%%%%%%%%%%%%%%%
\subsection{Channel Model and Received Signal}
%%%%%%%%%%%%%%%%%%%%%%%%%%%%%%%%%%%%%%%%%%%%%%%%%%%%%%%%%%%%%%%%
\label{sec:received_tag_signal}
\label{sec:channel_modelling}
Let $y_p(t)$ be the signal corresponding to a single tag reply where the effect of the channel between tag $p$ and the reader is captured.
Assuming a \LTI channel with flat fading during the short period of communication:
\begin{align}
y_p(t) = H_{RTR,p}\ A\ T_b\ c_p(t),
\label{eqn:y_p}
\end{align}
where $c_p(t)$ is the on--off key modulated square wave control signal for tag $p$ from \eqnref{eqn:c_p},
$T_b$ is the fraction of the power, which the tag is able to backscatter \cite{RFinRFID},
and $A$ is the amplitude of the transmitted carrier wave from the reader.
$H_{RTR,p}$ is the complex channel coefficient
$H_{RTR,p}=H_{RT,p}^2=H_{TR,p}^2$, which models that the channel coefficient between the reader and tag $p$, $H_{RT,p}$, is the same as the channel between tag $p$ and the reader, $H_{TR,p}$, due to reciprocity.
$H_{RTR,p}$ captures fading, antenna gains, and path--loss.
Then the received signal at the reader is:
\begin{align}
	z'(t) = \sum_{p=0}^{P-1}y_p(t-\tau_p) + L + O(t),
\label{eqn:z_hp}
\end{align}
where $P$ is the number of tags that participate in the response, $\tau_p$ is the unknown random delay for tag $p$, $L$ is the leakage from the reader's transmit antenna and the scatteres of the unmodulated carrier wave,
and $O(t)$ is \AWGN added at the reader.
More specifically, the antenna leakage can be decomposed as $L = H_{RR}A + L_{ant}$, where $L_{ant}$ is the antenna leakage, and $H_{RR}$ is the complex channel coefficient for the reader--to--reader channel.
We model $H_{RR,p}$ with Rayleigh fading because the line of sight component is captured by the direct antenna leakage $L_{ant}$.

Note that $y_p(t)$ in \eqnref{eqn:y_p} has infinite bandwidth because of the on--off keyed control signal $c_p(t)$.
However all other components in $y_p(t)$ are low--pass.
To model the effect of the receive filter at the reader, we introduce an ideal low--pass filter at the reader side, defined as $h_l(t) = 2W\sinc(2Wt)$,
where $W$ is the positive bandwidth of the low--pass equivalent signal.
Additionally, all tag replies are amplitude modulated,
thus only the envelope of the received signal contributes to the information in $z'(t)$.
The received low--pass envelope on the reader is:
\begin{align}
\nonumber z(t) &= \left|\int_{-\infty}^\infty z'(\tau)h_l(t-\tau)d\tau\right|\\
\label{eqn:z} &=\left|\int_{-\infty}^\infty\left[\sum_{p=0}^{P-1}H_{RTR,p}AT_bc_p(\tau-\tau_p) + H_{RR}A + L_{ant} + O(\tau)\right]h_l(t-\tau)d\tau\right|
\end{align}

%%%%%%%%%%%%%%%%%%%%%%%%%%%%%%%
\section{Parameter Set Estimation}
%%%%%%%%%%%%%%%%%%%%%%%%%%%%%%%
\label{sec:estimator}
%\begin{itemize}
%  \item Correlation/Wavelet framework (Should we weed out everything wavelet?)
%  \item Define mother and daughter wavelets/functions
%  \item Verification of framework (not as extensive as in the report)
%\end{itemize}
After having developed a detailed model for a tag signal and of its output through the communication channel,
the next step is to derive the structure for the signal parameters estimation, as shown in \figref{fig:overview}.
Specifically, the estimation module described in this section estimates the link frequency, $T_{l,p}$ and delay, $\tau_p$, 
of the strongest tag in the received signal, as defined in \eqnref{eqn:z}.

The information in a tag reply is encoded using the tag dependent control signal $c_p(t)$, 
which is true for all tags in the reply.
For estimating the two parameters, link frequency and delay, it is important to have a--priori known information about the structure of the tag replies in $z(t)$.
The estimation procedure is designed to exploit the fact that in a reply \emph{all} tags,
independent of the link frequency and delay chosen by tag $p$, \emph{use the same structure in the preamble control signal} $c_{pr,p}(t)$ to control the absorb and reflect state during backscatter of the preamble in a reply.
The structure in the preamble is the key used in the estimation framework introduced,
where a \emph{mother function}, $\psi(t)$, representing the preamble structure is designed.
Then, a number of derivatives of this mother function is created, denoted as \emph{daughter functions}, $\psi_{a,b}(t)$,
which are \emph{scaled} ($a$) and \emph{translated} ($b$) versions of the mother function.
\footnote{This framework is similar to that used in wavelet signal processing, see e.g. \cite{addison}.}
A daughter function is defined as:
\begin{align}
  \psi_{a,b}(t) = \psi\left(\frac{t-b}{a}\right)
\end{align}
Each daughter function is \emph{correlated} with the received signal $z(t)$ and the largest 
magnitude is used to estimate the frequency and delay of the strongest tag in the incoming signal.
This approach is motivated by the fact that the mean of the received preamble signal may be approximated by:
\begin{align}
  E[z(t)]=\alpha\psi\left(\frac{t-b}{a}\right)+\beta,\qquad b<t\leq b+D_{pr},
\end{align}
where $\alpha$ is an estimate of the signal level, $\beta$ is the offset added to remove the zero mean property of the mother function (more about this later), and $D_{pr}$ is the duration of the preamble.
The expectancy operation averages across the white noise. 

The correlation framework is defined using the received signal and a daughter function:
\begin{align}
T(a,b) = \left<z(t),\psi_{a,b}(t)\right> = \int_{-\infty}^{\infty} z(t)\psi_{a,b}(t) dt,
\label{eqn:cwt}
\end{align}
Calculation of $T(a,b)$ for a range of $a$ and $b$ results in a three--dimensional representation,
where a measure of the correlation of the received signal $z(t)$ with various daughter functions are given. Similarly, if the scalogram $E(a,b)=T^2(a,b)$ is considered,
then:
\begin{align}
(a_p,b_p)=\arg\max_{\substack{a\in\mathcal{A},\ b\in\mathcal{B}}}E(a,b)
\label{eqn:max_scalogram}
\end{align}
is the pair telling that it is very likely that tag $p$, with link frequency $f_{l,p}=\frac{1}{a_p}$, delayed $b_p$ is present in $z(t)$.
We use scalogram since if the channel has incurred a phase shift, the received signal has a large negative peak in the correlation representation.
The search ranges for $a$ and $b$, $\mathcal{A}$ and $\mathcal{B}$, respectively, are defined in the standard and depend on the settings of the reader.

The objective is to explicitly evaluate \eqnref{eqn:max_scalogram}.
%The search range defining the region where the problem in \eqnref{eqn:max_scalogram} is evaluated follows in \sectionref{sec:AB_search_range}.
The mother function is designed to capture the preamble structure in \sectionref{sec:mother_wavelet},
and the scaling and translation of the mother function leads to the definition of the daughter function in \sectionref{sec:daughter_wavelets}.

\subsection{Mother Function}
%%%%%%%%%%%%%%%%%%%%%%%%%%%%
\label{sec:mother_wavelet}
Let $\psi(t)$ be the real valued mother function that satisfies the following two requirements:
\begin{itemize}
  \item $\psi(t)$ must have finite energy, i.e. $\int_{-\infty}^{\infty} \psi^2(t) dt < \infty$.
    This ensures that the correlation is bounded in time.
  \item $\psi(t)$  must have no zero--frequency (DC) component in its support duration, i.e. it must be zero mean.
    Thereby the function is able to differentiate between signals based on their structure rather than their signal level.
\end{itemize}
%A consequence of the second property is that the reader leak included in $z(t)$ does not necessarily have to be removed prior to correlation with a daughter function.
Furthermore, the mother function is designed such that each bit in a symbol has duration $\frac{1}{2}\unit{s}$ ensuring that the link frequency of the mother function is $1\unit{Hz}$,
and so each symbol in the preamble has unit energy.
Thanks to this normalization the link frequency of a daughter function (which is designed in the following section) becomes $1/a$ when evaluated with the scaling parameter $a$.
The preamble structure consists of linear combinations of the basis functions derived in \secref{sec:sig_chan_model},
and the signal waveforms with unit energy are:
\begin{align}
\nonumber 																	s_0^{T=1}(t)&=\phi_0^{T=1}(t), 	& s_1^{T=1}(t)&=\phi_1^{T=1}(t),\\
\label{eqn:signal_const_mw_not_vectorized} 	s_2^{T=1}(t)&=-\phi_0^{T=1}(t), &	s_3^{T=1}(t)&=-\phi_1^{T=1}(t).
\end{align}
The mother function depends on whether FM0 or Miller is used.
Additionally, in the query sent by the reader,
the parameter \TRext{} specifies, which of two different preambles to use for a given encoding when a tag replies.
Let $N_{pr}$ be the number of symbols in a preamble,
and a $4\times N_{pr}$ state select matrix $\vS_{pr}$ is generated in the same way as in \secref{sec:sig_chan_model} from the preamble structures in \EPCglobal as:
\begin{align*}
\vS_{pr}^{FM0,TRext=0}&=
\begin{bmatrix}
\ve_2&\ve_3&\ve_4&\ve_1&\ve_4&\ve_2
\end{bmatrix}\\
\vS_{pr}^{FM0,TRext=1}&=
\begin{bmatrix}
\ve_1&\ve_1&\cdots&\ve_1&\vS_{pr}^{FM0,TRext=0}
\end{bmatrix}\\
&\hskip0.7cm\raisebox{0.7cm}{$\underbrace{\rule{2.9cm}{0cm}}_{12}$}
\vspace{-12pt}
\end{align*}
\begin{align*}
\vS_{pr}^{Miller,TRext=0}&=
\begin{bmatrix}
\ve_1&\ve_1&\ve_1&\ve_1&\ve_1&\ve_2&\ve_3&\ve_4&\ve_2&\ve_4
\end{bmatrix}\\
\vS_{pr}^{Miller,TRext=1}&=
\begin{bmatrix}
\ve_1&\ve_1&\cdots&\ve_1&\vS_{pr}^{Miller,TRext=0}
\end{bmatrix},\\
&\hskip0.7cm\raisebox{0.7cm}{$\underbrace{\rule{2.9cm}{0cm}}_{16}$}
\end{align*}
where $\ve_k$ indicates the state $s_{k-1}$, or equivalently that signal waveform $s_{k-1}(t)$ in \eqnref{eqn:signal_const_mw_not_vectorized} is used to generate the mother function.
With this in mind, let the square--wave modulated mother function be:
\begin{align}
\label{eqn:psit_prime}\psi(t)&=\sum_{n=0}^{N_{pr}-1}\sum_{k=0}^3\ve_{k+1}^\transpose\vS_{pr}\ve_{n+1}s_k^{T=1}(t-nM)
=\sum_{n=0}^{N_{pr}-1}\sum_{k=0}^1\ve_{k+1}^\transpose\vV\vS_{pr}\ve_{n+1}\phi_k^{T=1}(t-nM),
\end{align}
where $\vV$ is the signal constellation matrix from \eqnref{eqn:sig_space_rep}.
Notice that the preamble structure has zero mean and that the inherent memory structure of both FM0 and Miller encoding is violated in the preambles.
This ensures that it is not possible for the designed function to correlate as strongly to the data as to the preamble.

A final consideration on the design is that $z(t)$ contains the low-passed filtered signal.
However, since each daughter function is recalculated for each value of $a$ and $b$,
to contain the computational complexity we consider the daughter functions square-wave modulated.

%%%%%%%%%%%%%%%%%%%%%%%%%%%
\subsection{Daughter Function}
%%%%%%%%%%%%%%%%%%%%%%%%%%%
\label{sec:daughter_wavelets}
A daughter function of the mother function is defined as:
\begin{align}
\psi_{a,b}(t) = w(a)\psi\left(\frac{t-b}{a}\right),
\label{eqn:dwl}
\end{align}
where the scaling and translation parameters are used in the evaluation of the mother function,
and where $w(a)$ is a weighti, which ensures that all configurations of a daughter function are equally weighted and not biased by the parameter set $(a,b)$ when matched onto the input signal in the correlation $\left<z(t),\psi_{a,b}(t)\right>$.
The amplitude level for a tag when a bit is backscattered is the same during communication no matter what link frequency is used.
Thus the energy in a tag preamble oscillating with a fast frequency is less than the energy in a tag preamble oscillating with a slow frequency.
The match constraint to determine $w(a)$ therefore takes into account $z(t)$.

\begin{lem}
\label{lem:wa}
(Proved in the Appendix) The weight ensuring that daughter functions are correctly scaled for all values of $a$ and $b$ in a reply containing only one tag reply is:
\begin{align}
w(a)=\frac{1}{a}.
\label{eqn:w_a}
\end{align}
\end{lem}

%%%%%%%%%%%%%%%%%%%%%%%
\section{Data Decoding}
%%%%%%%%%%%%%%%%%%%%%%%
\label{sec:data_decoding}
%\begin{itemize}
%  \item Matching Pursuit
%  \item Channel attenuation ($\alpha$) estimation
%  \item Viterbi as MLSD (I will not explain Viterbi as thoroughly as in the report)
%  \item Multiple tag decoding as part of the Q-protocol
%\end{itemize}
The estimates of the link frequency $\frac{1}{a}$ and the delay offset $b$ obtained using the framework presented in the above makes it possible to decode one and possibly several tag replies in a received signal,
even when they are dispersed in time and frequency.
This corresponds to the tag resolution part shown in \figref{fig:overview} and is treated in this section.

One shortcoming of using UHF RFID as a use case for general multiple sensor decoding is that \EPCglobal does not currently support multiple Ack commands after a collided reply, 
which clearly affects the possible gain in data decoding.
In the description of the data decoding algorithm that follows, it is assumed that multiple Ack commands are allowed in the standard, to show the true potential of our method.

To decode multiple tag replies an iterative, greedy algorithm, \SIC, is used.
Before explaining the \SIC algorithm further, 
the next section derives the optimum decoder structure for a single tag, using the above described estimators.

%%%%%%%%%%%%%%%%%%%%%%%%%%%%%%%%%%%%%5
\subsection{Optimized Single Tag Decoding}
%%%%%%%%%%%%%%%%%%%%%%%%%%%%%%%%%%%%%5
\label{sec:optimized_single_tag_decoding}
The optimal algorithm for a detector, where the memory in a sequence satisfies the Markov property as the $n$th symbol in a sequence \emph{only} depends on the $(n-1)$th decoded symbol, is the Viterbi algorithm \cite{proakis}.
In addition to the memory structure in the encoding scheme,
there are two additional a--priori known structures in \EPCglobal to improve the decoding:
\begin{enumerate}
  \item The last symbol in the preamble and the signal waveform used to create it is a--priori known and are previously found in \eqnref{eqn:s_init}.
  \item The postamble symbol after the data part is a--priori known to be symbol--1.
\end{enumerate}
Recall from \sectionref{sec:control_signal} that the structure containing \emph{both} the memory, the data message $\mathcal{M}$ and the postamble symbol is the 4--by--$(N_{\mathcal{M}}+1)$ state select matrix $\vS=\begin{bmatrix}\vs_1&\vs_2&\cdots&&\vs_{N_{\mathcal{M}}+1}\end{bmatrix}$, where each $\vs_i$ is a coordinate vector $\ve_k$ and $k=1,2,3,4$ indicates which of the respective signal waveforms $s_0(t)$, $s_1(t)$, $s_2(t)$, and $s_3(t)$ is used to encode a symbol in the tag reply.
The objective is therefore reformulated to estimate the state select matrix $\vS$ as it contains the memory structure, encoded message, and postamble.

Let the signal processed in the $i$th iteration be $r_i(t)$, with $r_0(t)=z(t)$, and assume that it only contains one tag reply with parameters $a$ and $b$.
The expected value of $r_i(t)$ is then similar to the control signal $c_p(t)$ in \eqnref{eqn:c_p}:
\begin{align}
  \label{eqn:z_approx} E[r_i(t)] &= \alpha_i\Big(\psi^1_{a_i,b_i}(t) + \sum_{n=0}^{N_{\mathcal{M}}}\sum_{k=0}^3 \ve_{k+1}^\transpose\vS\ve_{n+1}s_k^{a_i}\left(t-b_i-D_{pr,i}-na_iM\right) + \gamma_i(t)\Big) + \beta,
\end{align}
where $\alpha_i$ is the signal level at the reader side.
Notice here that the daughter function has a superscript $1$ attached.
This is because this daughter function must be scaled to have \emph{unit energy per symbol},
rather than to have correct scaling for all values of $a$ and $b$ in a reply containing only one tag signal.
This results in changing the weight in \eqnref{eqn:w_a} from $\frac{1}{a}$ to $\frac{1}{\sqrt{aM}}$.
This change in scaling allows for the correct scaling afterwards to the signal level, $\alpha_i$.
$\gamma_i(t)$ is the offset introduced to ensure signal levels that correspond to the way a tag backscatters its reply (recall \eqnref{eqn:signal_enc}), 
$\beta$ is an estimate of the reader leak and $D_{pr,i}=a_iMN_{pr}$ is the estimated duration of the preamble.
The signal waveforms:
\begin{align}
\label{eqn:stildedecoder} s_0^{a_i}(t)&=\phi_0^{a_i}(t) & s_1^{a_i}(t)&=\phi_1^{a_i}(t) & s_2^{a_i}(t)&=-\phi_0^{a_i}(t) & s_3^{a_i}(t)&=-\phi_1^{a_i}(t),
\end{align}
follow from the symbol basis functions in \eqnref{eqn:phiFM0} and \eqnref{eqn:phiMiller} and can be represented in terms of the signal space representation matrix $\vV$ as:
\begin{align}
  s_k^{a_i}(t) = \sum_{j=0}^1 \ve_{j+1}^\transpose\vV\ve_{k+1}\phi_j^{a_i}(t).
  \label{eqn:stildedecoder_generel}
\end{align}

\begin{figure}[h]
\centering
\begin{pspicture}(-0.5,0)(4.5,2)
\psframe[framearc=0.3](1,0)(3,2)
\psline{->}(0,0.4)(1,0.4)
\psline{->}(0,0.8)(1,0.8)
\psline{->}(0,1.2)(1,1.2)
\psline{->}(0,1.6)(1,1.6)
\psline{->}(3,1)(4,1)
\uput[180](0,1.6){initial state}
\uput[180](0,1.2){end state candidates}
\uput[180](0,0.8){$\vZ$}
\uput[180](0,0.4){$\vH$}
\uput[0](4,1){$\vSh$}
\rput(2,1){MLSD}
\end{pspicture}
\caption{Black box illustration of a MLSD. A correlation matrix $\vZ$ and a transition matrix $\vH$ are used with the information of the initial state and possible end states to estimate the most likely transmitted state select matrix $\vSh$.}
\label{fig:viterbi_black_box}
\end{figure}
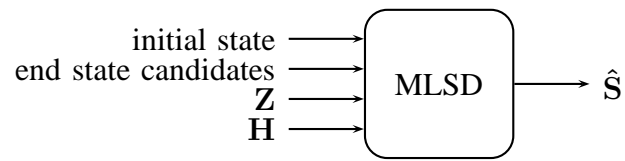

Let the outcome of the \MLSD be an estimate of $\vS$, denoted $\vSh$, and consider the Viterbi algorithm as the black box for \MLSD in \figureref{fig:viterbi_black_box};
for optimal decoding it uses the three data structures:
\begin{itemize}
	\item Initial state vector $\vs_{\text{init}}$ --- The initial state seeding the decoding which follows from the state corresponding to the last symbol in the preamble. For FM0, this state is $s_1$ and for Miller it is $s_3$,
and the vector is defined in \eqnref{eqn:s_init}.
	\item A $4\times(N_{\mathcal{M}}+1)$ matrix $\vZ$ --- Cost matrix not considering memory, where each element represents how well each of the four signal waveforms $\stilde_k(t)$, $k=0,1,2,3$,
	match a singled out symbol part in the residual, where a large element value indicates a good match.
	Values are found for the $N_{\mathcal{M}}=16$ data symbols and the postamble symbol.
	\item A $4\times4$ matrix $\vH$ --- Memory structure representing the allowed paths between states observed for two adjacent encoded symbols.
\end{itemize}
The memory structure matrix $\vH$ is given by the valid state transitions.
In the generation of the control signal at the tag in \sectionref{sec:control_signal},
the two matrices $\vH_{m_n=0}$ and $\vH_{m_n=1}$ are derived describing the transition to make, conditioned on the symbol $m_n$ to send.
$\vH$ follows as the version where the transmitted symbol is unknown:
\begin{align*}
  \vH^{FM0} = \vH_0^{FM0} + \vH_1^{FM0} \qquad\qquad \vH^{Miller} = \vH_0^{Miller} + \vH_1^{Miller}
\end{align*}
and $\vH_{k,k'}=1$ indicates that it is possible to go from state $s_{k'-1}$ to state $s_{k-1}$.

The two matrices $\vZ$ and $\vH$ can be represented as a trellis, 
where each node (except the node in the preamble) represent the entries of $\vZ$ and the possible transitions are given by the entries in $\vH$.
The reason why there are only two possible states for the first symbol is that the previous state is known a--priori.
The same applies for the postamble as it is known to be a symbol--1.

It is useful to remark that, if the channel for the tag to be decoded in the $i$th iteration incurs a complete phase shift on the backscattered reply in the residual,
this means that the state of the last symbol in the preamble is $s_3$ instead of $s_1$ for FM0, and $s_1$ instead of $s_3$ for Miller.
The phase shift can be detected in several ways; for example if $\alpha_i<0$ or $T_i(a_i,b_i)<0$ a phase shift is introduced.
In the sequel this effect is neglected to simplify the description of the decoding,
however, it is important for an implementation to detect the phase shift and flip the initializing state in the decoder.

%%%%%%%%%%%%%%%%%%%%%%%%%%%%%%%%%%%%%%%%%%%%%%%%%%%%%%%%%%%%%%%%%%%
\subsection{Successive Interference Cancellation for Data Decoding}
\label{sec:mp_tag_decoding}
%%%%%%%%%%%%%%%%%%%%%%%%%%%%%%%%%%%%%%%%%%%%%%%%%%%%%%%%%%%%%%%%%%%
The \SIC algorithm works by iteratively estimating the strongest signal component in a signal consisting of multiple signal components.
The estimated signal component is then subtracted from the others, and the next most strongest signal component is estimated and subtracted in the following iteration.
The $i$th iteration of the algorithm is defined as follows:
\begin{align*}
  r_{i+1}(t) = r_{i}(t) - q(t),
\end{align*}
where $r_{i}(t)$ is the current residual signal, $q(t)$ is an estimate of the strongest signal component in $r_{i}(t)$ and $r_{i+1}(t)$ is the resulting residual, used in the next iteration.

The $i$th estimate of the contribution to subtract from the residual follow from \eqnref{eqn:z_approx} where the estimate of the transmitted message $\vSh$ is used to model the tag contribution signal $\qh_i(t)$:
\begin{align}
\label{eqn:qh_i_data_decoding}\qh_i(t) &= \alpha_i\Big(\psi^1_{a_i,b_i}(t) + \sum_{n=0}^{N_{\mathcal{M}}}\sum_{k=0}^3 \ve_{k+1}^\transpose\vSh\ve_{n+1}\cdot
s_k^{a_i}\left(t-b_i-D_{pr,i}-na_iM\right) + \gamma_i(t)\Big),
\end{align}
where the $i$th modulation depth estimate $\alpha_i$ follow from \lemmaref{lem:alpha} in the Appendices and $\gamma_i(t) = w^1(a_i)\rect\left(\frac{t-b_i-\frac{D_{pr,i}+a_i(N_{\mathcal{M}}+1)}{2}}{D_{pr,i}+a_i(N_{\mathcal{M}}+1)}\right)$.
The weight, $w^1(a_i)$, is to ensure unit energy per symbol of the signal before scaling with $\alpha_i$.

The algorithm is summarized as follows:
Let $\mathcal{H}=\{v_0,v_1,\ldots\},v_i=(a_i,b_i)$ be the history of estimates, assume $z(t)$ contains only one tag reply, set $r_0(t)=z(t)$ and let $i=0$, then:
\begin{enumerate}
	\item Find the $i$th scalogram, as in \eqnref{eqn:cwt}, and determine whether a phase shift has occurred.
	\item Find estimates for the strongest contribution in the newly found scalogram:
	\[ v_i=(a_i,b_i)=\argmax_{\substack{a\in\mathcal{A},\\b\in\mathcal{B}}}E_i(a,b) \]
	\item Decode a message from $r_i(t)$ with location parameters $(a_i,b_i)$ using the Viterbi algorithm described in the previous section.
%	\item Send an Ack command with the decoded message. %If the \EPC from the tag with the decoded tag token (\RN) is not returned then break.
	\item Generate an estimate of the complete tag contribution $\qh_i(t)$ from \eqnref{eqn:qh_i_data_decoding}.
	\item Subtract the estimate from the residual $r_{i+1}(t)=r_i(t)-\qh_i(t)$.
	\item Let $\mathcal{H}=\{\mathcal{H},v_i\}$ if $\mathcal{H}\cap v_i=\{\}$, increment $i$, and re--iterate.
\end{enumerate}
The termination criteria for the algorithm depends on the scenario it is used in and what happens when an Ack is received at a sensor.
In the case of UHF RFID tags, there are several possible methods, which are treated in the next section, 
which concerns the implementation of the $Q$--protocol, used in UHF RFID.

%%%%%%%%%%%%%%%%%%%%%%%%%%%%%%%%%%%%%%%%%%%%%%%%%%%%%%%%%%%%%%
\section{Implementing Multiple Tag Decoding in the $Q$--Protocol}
%%%%%%%%%%%%%%%%%%%%%%%%%%%%%%%%%%%%%%%%%%%%%%%%%%%%%%%%%%%%%%
\label{sec:q_protocol}
With the algorithm concluding the previous section, it is possible to decode a single slot shown in \figref{fig:overview}.
This section describes how this is extended to being part of an entire arbitration protocol run, using the $Q$--protocol of \EPCglobal.
It is useful to evaluate the effect multiple tag decoding has on the Q-protocol,
if acknowledging multiple tags is allowed.
To enable this, the Q-protocol is implemented in MATLAB and Monte Carlo simulations are run,
to determine how much \emph{time} it takes to resolve an entire tag population and how many \emph{transmissions} it takes.
The results can be used to evaluate the following (1) Is multiple tag decoding in the $Q$--protocol more time efficient than single tag decoding?
(2) Is multiple tag decoding more energy efficient, with respect to transmissions count from the reader?

In the numerical simulations the transmissions by both reader and tag are counted as listed in \tabref{tab:reader_and_tag_commands}. Based on the duration of each of these commands and the three timeouts $T_1$, $T_2$ and $T_3$ from the standard,
the duration of the inventorying can be calculated.
\begin{table}
	\centering
	\begin{tabular}{|l|l|c|}\hline
		\multicolumn{3}{|l|}{\textbf{\underline{Reader}}}\\\hline
		\textbf{Name} & \textbf{Contents other than payload} & \textbf{No. of bits in payload} \\\hline
		Query & delimiter, data-0, RTcal, TRcal & 22 \\\hline
  	QueryRep & delimiter, data-0, RTcal & 5 \\\hline
  	Ack	& delimiter, data-0, RTcal & 18 \\\hline\hline
		\multicolumn{3}{|l|}{\textbf{\underline{Tag}}}\\\hline
		\textbf{Name} & \textbf{Contents other than payload} & \textbf{No. of bits in payload} \\\hline
		RN16 & preamble, postamble & 16 \\\hline
  	PC/XPC + EPC + CRC & preamble, postamble & 128 \\\hline
	\end{tabular}
	\caption{Transmitted reader and tag commands \cite{epc}.}
	\label{tab:reader_and_tag_commands}
\end{table}

%\begin{figure}
%\centering
%\subfloat[Reader model.]{\label{fig:reader_model}\includegraphics[scale=0.3]{Qprotocol/images/readermodel.eps}}
%\qquad
%\subfloat[Tag model.]{\label{fig:tag_model}\includegraphics[scale=0.3]{Qprotocol/images/tagmodel.eps}}
%\\
%\subfloat[Tag state diagram.]{\label{fig:tag_state_diagram}\includegraphics[scale=0.3]{Qprotocol/images/tagstatediagram.eps}}
%\caption{Reader and tag model used. A counter is maintained for each transmission type and used to calculate the arbitration time and transmission/energy efficiency afterwards.}
%\label{fig:readertagmodel}
%\end{figure}

\subsection*{Design Assumptions for Q--Protocol Implementation}
Prior to an experiment, we assume a Select command has been issued and received correctly by all tags.
This defines the scope of the experiment to inventorying of tags alone.
Additionally, all tags have their inventoried flag for the selected session set to the same value, either A or B.
This means that out of $N$ tags, $N$ tags participate in the inventorying.

For simplicity the command QueryAdjust is not used \emph{during} a round to change the value of $Q$.
Instead, a round is always completed with a chosen $Q$ after which a new Query command is issued with a new value of $Q$.
A QueryAdjust can increment or decrement $Q$ during an inventory round, and when to use it during arbitration must be analyzed thoroughly first,
to understand its effect on time and energy usage.
We have therefore not optimized the $Q$--protocol for multiple tag decoding and we expect that a higher gain is achievable if this is done.

It is assumed that the reader can determine perfectly whether a slot is idle or not and whether a slot contains any remaining tag signals.
This is in order to focus on tag decoding, not on detecting whether there are tags to be decoded.
In a future implementation, this detection could instead be done based on signal levels.
As the variance of the noise can be estimated before tag to reader communication,
it can be decided whether one or more tags are present, if enough samples cross a detection threshold based on the variance during tag to reader communication.
This threshold may also be used for detecting when the residual in the \SIC algorithm contains no more tags.
In the case of UHF RFID tags an Ack transmission from a reader is quite expensive and should be avoided if possible.
Otherwise, the termination criteria of the algorithm could be based on a \emph{digital decision},
where the \SIC algorithm terminates if no sensor replies the Ack.

Because each iteration of the \SIC algorithm is dependent on the previous iterations and the accuracy of the estimates of $a$, $b$ and the signal level,
the estimation of the signal level is assumed perfect in this implementation, to focus exclusively on the impact of multiple tag decoding based on estimation of link frequency, $a$, and delay, $b$.
Also, the reader is assumed to be unable of detecting a collision, it will always attempt to decode the strongest tag.
If a tag is correctly singled out, the Ack and EPC are correctly transferred and decoded.
This, to allow for simplicity and because if a tag RN16 has already been successfully singled out, the probability of error in receiving the Ack and transmitting the EPC is smaller.
Additionally, we assume that the forward link (Reader-to-Tag) is error free, to be able to focus fully on multiple tag decoding.

When a single tag is resolved, an extra frame is conducted, to ensure that no weaker tags are unresolved.
For fair comparison, this is done for both the original and new reader.
A change in the UHF RFID standard for the tags is assumed, namely that when a tag receives an Ack with a \emph{wrong} RN16, it does not transition to state arbitrate, but remains in state reply.
Only when a Query or QueryRep command is received does the tag transition to state arbitrate. 
This allows for multiple tag acknowledging by sending an Ack and receiving and decoding the EPC of the resolved tag in each iteration of the \SIC,
rather than only being able to send on Ack per slot.

%%%%%%%%%%%%%%%%%
\section{Results}
%%%%%%%%%%%%%%%%%
\label{sec:results}
%\begin{itemize}
%  \item Figure 10.5b from the report
%  \item Numerical evaluation of Q protocol with multiple tag decoding - how to choose Q
%  \item Gain/Throughput evaluation
%\end{itemize}
To show the benefit of having multiple tag decoding in a reader,
we have performed two simulation tests.
The first test illustrates the probability of decoding a given number of tags.
The second test is a comparison of the duration of an inventory round using the $Q$--protocol when using a reader with and without multiple tag decoding.
The tests have been made for a scenario where the tags choose their link frequency according to a Gaussian distribution 
with mean equal to a nominal \BLF of $50\unit{kHz}$ and a variance such that $99.73\%$ ($3\sigma$ using the empirical rule) of the generated link frequencies and delays fall within the \EPCglobal requirements.
Also, a long preamble ($TRext=1$), $FM0$ encoding is used, 
the distance between reader and tags is set to $1$ meter and 
the low--pass filter employed has a bandwidth of $1.5\unit{MHz}$.
The noise power at the reader antenna is set to $-50\unit{dBm}$.

\begin{figure}
\centering
\includegraphics[width=0.6\textwidth]{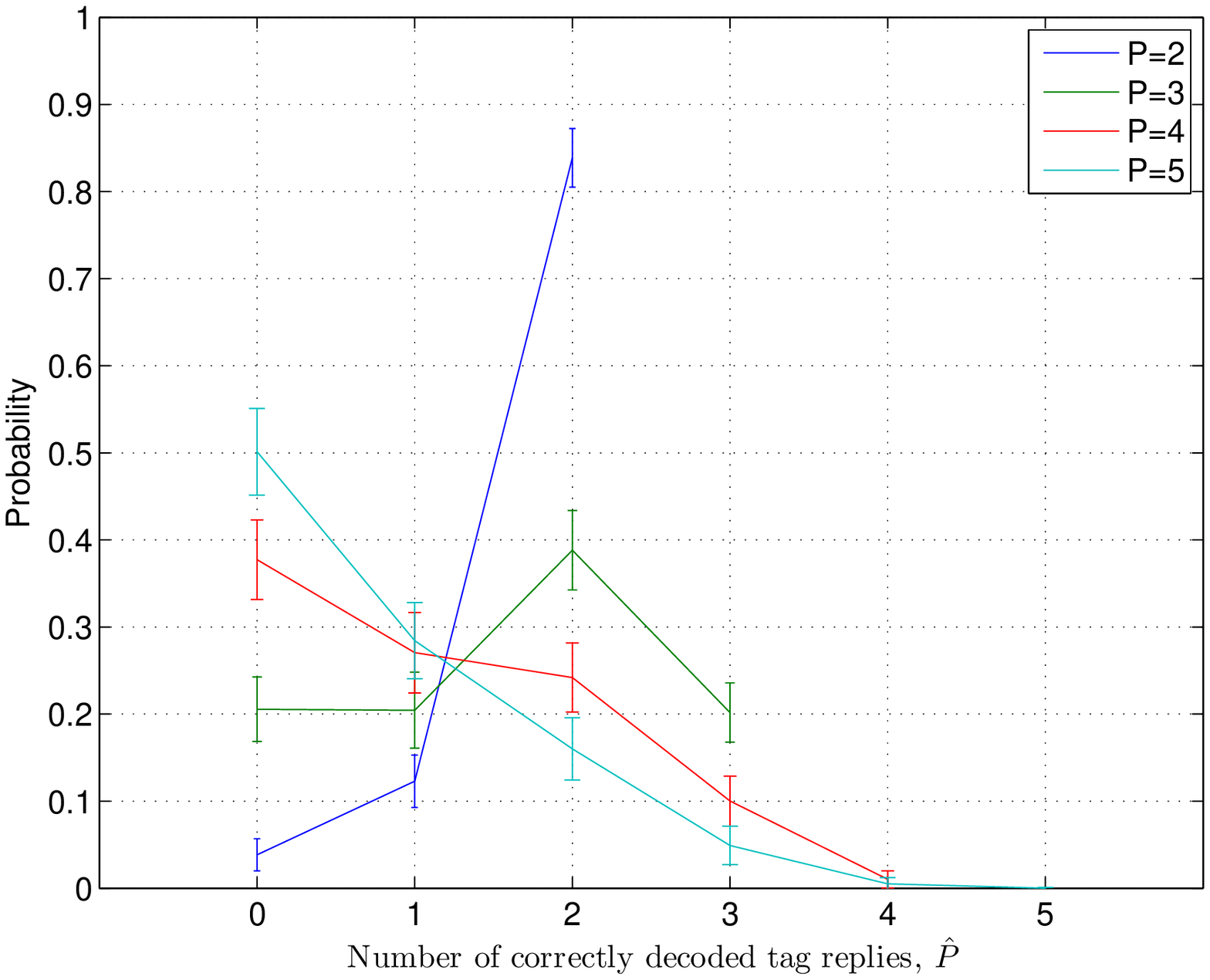}
\caption{Numerically simulated collision resolution success probability for multiple tag decoding. The distance is 1 meter and $P=\{2,3,4,5\}$ tag replies are collided.}
\label{fig:distribution_of_number_of_resolved_tags}
\end{figure}

In the first test a number of experiments and runs are performed.
A \emph{run} is defined as the generation of a received signal, as in \eqnref{eqn:z}, and the following decoding of that signal.
After decoding, the estimated message and the actual encoded message is compared, and if the decoding was incorrect, the run is marked as \emph{erroneous}.
An \emph{experiment} is a series of runs.
For the first test, 100 experiments, each containing 100 runs, has been performed.
The results are shown in \figref{fig:distribution_of_number_of_resolved_tags},
where the gain for four different tag cardinalities, $P=\{2,3,4,5\}$ is illustrated.
The result of one experiment is used to calculate the percentage of runs ending in a given number of decoded tags.
The results from all the experiments are then used to calculate the standard deviation of this statistic.
The figure shows that it is possible to decode multiple tags, even when there are up to five tags present in the collision.
Even though the total five tags are rarely decoded,
the results show that in $50\%$ of the cases some of the tags are decoded,
which is a gain compared to presently used methods.

\begin{figure}
\centering
\includegraphics[width=0.6\textwidth]{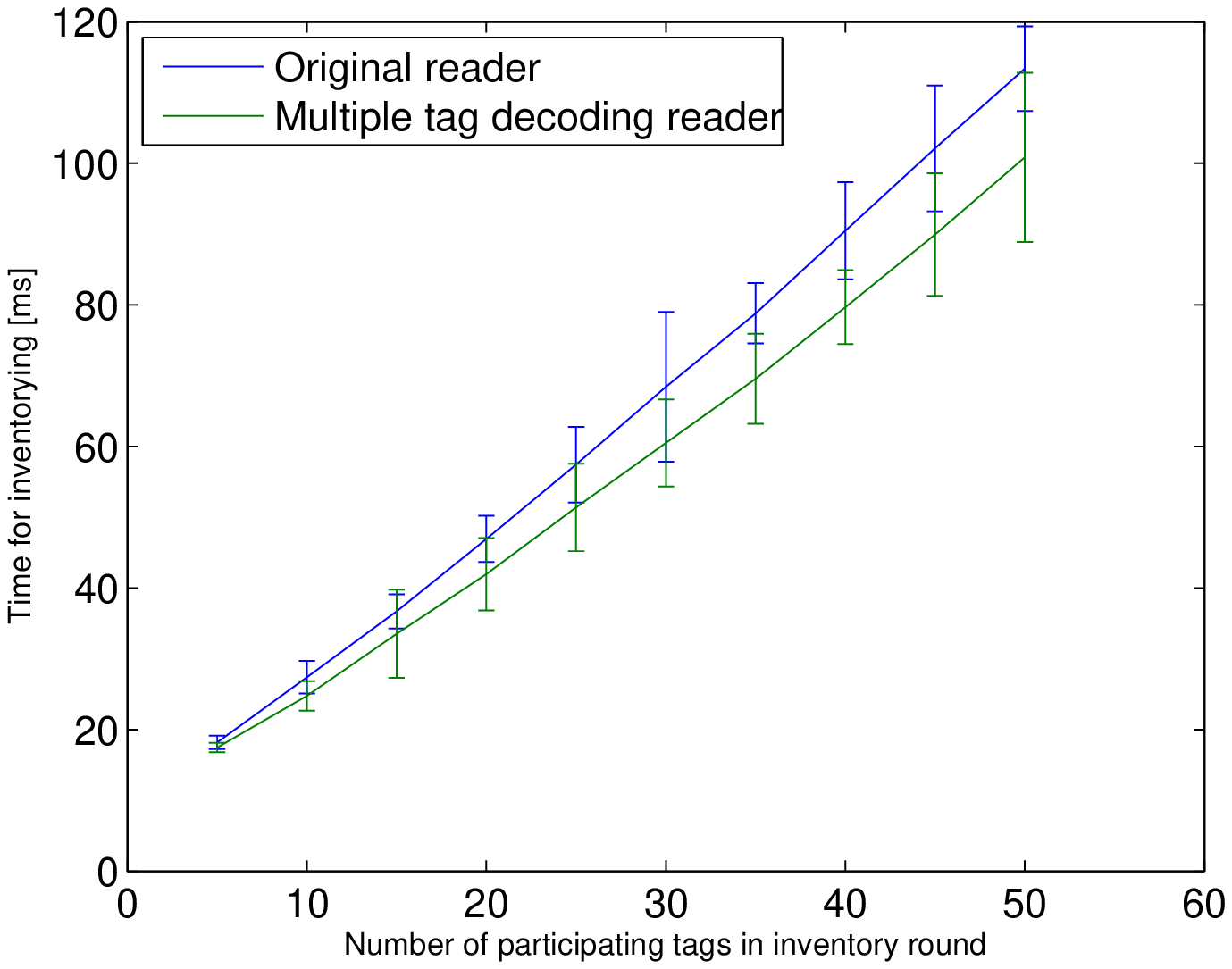}
\caption{The duration of the inventorying in the numerical simulation of multiple tag decoding in the $Q$--protocol.}
\label{fig:durations}
\end{figure}

\begin{figure}
\centering
\includegraphics[width=0.6\textwidth]{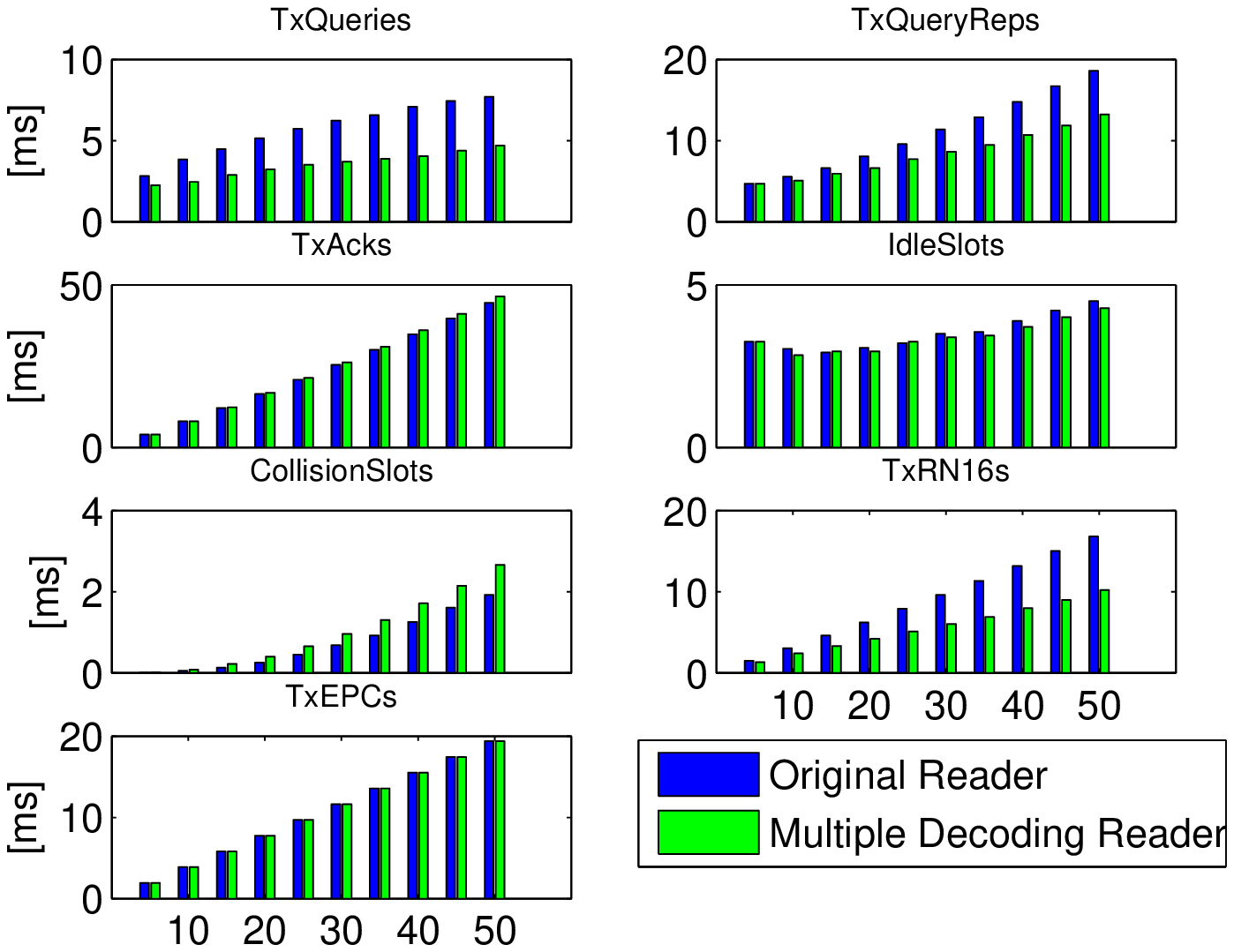}
\caption{The distribution of the transmitted commands during inventorying in the numerical simulation of multiple tag decoding in the $Q$--protocol.}
\label{fig:timeUsage}
\end{figure}

In the other test, we implement the $Q$--protocol with multiple tag decoding and compare it to 
a normal reader, which only decodes one tag per slot.
The initial value of $Q$ is set to $4$ and 1000 runs are conducted.
In each run, a randomly generated tag set is resolved using the Q protocol.
The time it takes to resolve the tag set is found, by counting the transmitted commands, as specified in the previous section.
The result is averaged over the 1000 runs and plotted in \figref{fig:durations}.
The distribution of the commands is further elaborated on in \figref{fig:timeUsage}.
As can be seen from \figref{fig:durations}, multiple tag decoding decreases the duration of the inventorying, especially for a large number of tags. 
From \figref{fig:timeUsage}, it is clear that with multiple tag decoding, fewer Queries and QueryReps are sent.
The RN16 count must be further explained.
If more than one tag transmit their RN16 at the same time, there is a collision,
which is counted as one RN16, as the duration is independent of the number of participating tags in the collision.
The number of RN16s sent out has also decreased dramatically, as several tags can be decoded in one slot.
The number of collisions has increased when using multiple tag decoding, but by a very small amount when compared to the savings.
The number of idle slots and acknowledgements are roughly the same for both single and multiple tag decoding.
The number of EPC is exactly identical, as is expected for full resolution.
The reason why the difference is not larger in \figref{fig:durations} is that the number of the most expensive transmissions, the acknowledgement, is unchanged. 
If the protocol is changed to allow for acknowledging multiple tags with a single composite Ack, the performance will greatly improve. 
Overall, the results show that multiple tag decoding does provide savings in time and energy and this gain increases approximately linearly, 
meaning that for tag populations in the hundreds and thousands,
this would provide a significant increase in time and power efficiency.

%%%%%%%%%%%%%%%%%%%%
\section{Conclusion}
%%%%%%%%%%%%%%%%%%%%
\label{sec:conclusion}
%\begin{itemize}
%  \item Tag variation may be viewed as a possibility rather than a problem
%  \item MLSD allows for multiple tag decoding together with good estimates of frequency and delay
%  \item Multiple tag decoding allows for much faster resolution of tag sets, when incorporated into the Q protocol
%\end{itemize}
The concepts presented in this paper show that tag variability can be transformed from foe to friend,
by using such differences to decode multiple colliding UHF RFID tag replies.
By utilizing the knowledge of the tag signal, it is shown to be feasible to distinguish between individual tag signals in numerical simulations.
Also presented is a detailed mathematical model of the tag signals using standard signal representation techniques,
which, to the knowledge of the authors, has not been presented in this level of detail before for both FM0 and Miller encoding.
The final tests where multiple tag decoding is incorporated into the $Q$--protocol shows a potential for time and transmission savings,
in terms of fewer transmitted commands from the tags.
%This is done under some ideal assumptions, which must be challenged before final conclusions may be drawn, 
%but these initial results show that the idea is feasible.

\appendices
\section{Proof of Weight Ensuring Correct Scaling of Daughter Functions}
\begin{lem}
\label{lem:wa}
The weight ensuring that daughter functions are correctly scaled for all values of $a$ and $b$ in a reply containing only one tag reply is:
\begin{align}
w(a)=\frac{1}{a}.
\label{eqn:w_a}
\end{align}
\end{lem}
\begin{IEEEproof}
The following simplifications are made for the derivation:
\begin{itemize}
	\item Only one tag is assumed to be present in the reply $z(t)$ with link frequency $\frac{1}{a}$,
	and the duration of the encoded preamble in the reply is $D_{pr}$, 
    that is, the tag preamble contributes to $z(t)$ for $b<t<b+D_{pr}$.
	\item $w(a)$ is determined for the pair $a, b$ that leads to a maximum or minimum in \eqnref{eqn:cwt},
	i.~e. only the case where the duty cycle duration $a$ for the encoded tag in $z(t)$ and the duration of the mother function $MN_{pr}$ satisfy $D_{pr}\equiv aMN$, where $M$ is the number of subcarrier cycles per symbol and $N_{pr}$ is the number of symbols in the preamble.
\end{itemize}
The property to be satisfied is that a daughter function, when correlated with $z(t)$ should satisfy a parameter independent correlation level:
\begin{align}
\left<E[z_1(t)],\psi_{a_1,b_1}(t)\right>=\left<E[z_2(t)],\psi_{a_2,b_2}(t)\right>
\label{eqn:wa_problem}
\end{align}
should be satisfied, where the tag reply in $z_j(t)$ is encoded with \emph{different} link frequency $\frac{1}{a_j}$ and delay $b_j$ but with the \emph{same channel},
thus the signal levels in $z_1(t)$ and $z_2(t)$ are equal.
In the interval $b<t<b+D_{pr}$ the received signal $z(t)$ has the property that its expected value can be written in terms of a weighted mother function with the same configuration as the control signal used to model $z(t)$:
\begin{align}
  E[z(t)]=\alpha\psi\left(\frac{t-b}{a}\right)+\beta,\qquad b<t<b+D_{pr},
\label{eqn:z_approx}
\end{align}
where $\alpha$ controls the signal level and $\beta$ the DC component in $z(t)$.
Rewriting, letting $t'=\frac{t-b}{a}$:
\begin{align}
E[z(at'+b)]=\alpha\psi(t')+\beta,\qquad 0<t'<MN_{pr}.
\label{eqn:ztb}
\end{align}
As the signal levels in $z(t)$ clearly does not depend on the encoding parameters $a$ and $b$,
then the correlation of \eqnref{eqn:ztb} with the mother function is therefore \emph{constant for all encoded tag replies with different $a$ and $b$}.
That is:
\begin{align}
\left<E[z_1(a_1t+b_1)],\psi(t)\right> = \left<E[z_2(a_2t+b_2)],\psi(t)\right>
\label{eqn:wa_solution}
\end{align}
is a property that is always satisfied when $z_j(\cdot)$ is encoded with parameters $a_j$ and $b_j$,
and thus is the property requested in \eqnref{eqn:wa_problem}.
As the daughter function is a scaled and translated version of the mother function,
combine \eqnref{eqn:wa_problem} and \eqnref{eqn:wa_solution}:
\begin{equation}
\int_{-\infty}^{\infty}E[z(at+b)]\psi(t)dt = \int_{-\infty}^{\infty}E[z(t)]w(a)\psi\left(\frac{t-b}{a}\right)dt =w(a)a\int_{-\infty}^{\infty}E[z(at'+b)]\psi(t')dt'
\nonumber
\end{equation}
where $t' = \frac{t-b}{a}$ and $dt' = \frac{dt}{a}$, and the weight is $w(a)=\frac{1}{a}$ which completes the proof.
\end{IEEEproof}

\section{Proof of Optimal Estimator of $\alpha$}
\begin{lem}
\label{lem:alpha}
Aim for the lowest contribution for the tag with parameter configuration $(a_i,b_i)$ in the scalogram evaluated in the next iteration $i+1$, then the optimal estimator for $\alpha_i$ is:
\begin{align}
\alpha_i=\frac{\sqrt{a_i}T_i(a_i,b_i)}{\sqrt{M}N_{pr}},
\end{align}
where $T_i(a_i,b_i)$ is the \CWT value for the estimated $a_i$ and $b_i$, $M$ is the number of subcarrier cycles per symbol in the encoding scheme and $N_{pr}$ is the number of symbols in the preamble.
\end{lem}
\begin{IEEEproof}
The problem to be optimized is:
\[ \alpha_i=\argmin_{\alpha\in\mathbb{R}}E_{i+1}(a_i,b_i), \]
i. e. find the $\alpha_i$ which minimizes the contribution in iteration $(i+1)$ of the tag found in iteration $i$.
The scalogram as a function of the $i$th residual and the contribution to be removed is:
\begin{align}
\nonumber &E_{i+1}(a_i,b_i) = \left(\int_{-\infty}^{\infty}E[r_{i+1}(t)]\psi_{a_i,b_i}(t)dt\right)^2 =\left(\int_{-\infty}^{\infty}\left[E[r_i(t)]-\qh_i(t)\right]\psi_{a_i,b_i}(t)dt\right)^2\\
\nonumber &\overset{i}{=}\left(\int_{-\infty}^{\infty}\left[E[r_i(t)]-\alpha_i(\psi^1_{a_i,b_i}(t)+\gamma_i(t))\right]\psi_{a_i,b_i}(t)dt\right)^2\\
\nonumber &=\left(T_i(a_i,b_i)-\alpha_i\int_{-\infty}^{\infty}\psi^1_{a_i,b_i}(t)\psi_{a_i,b_i}(t)dt - \alpha_i\int_{-\infty}^{\infty}\gamma_i(t)\psi_{a_i,b_i}(t)dt\right)^2\overset{ii}{=}\left(T_i(a_i,b_i)-\frac{\alpha_i\sqrt{M}N_{pr}}{\sqrt{a_i}}\right)^2\\
\label{eqn:alpha_quadratic_function}&=\alpha_i^2\frac{MN_{pr}^2}{a_i}-\alpha_i\frac{2\sqrt{M}N_{pr}T_i(a_i,b_i)}{\sqrt{a_i}}+E_i(a_i,b_i),
\end{align}
where $i)$ follows as the support duration of $\psi_{a_i,b_i}(t)$ ensures that $\qh_i(t)$ is not evaluated in the part where the extra half symbol is added to $\qh_i(t)$,
and where $ii)$ follows firstly because the daughter function is zero mean in the interval of the support duration of $\gamma_i(t)$ and secondly by evaluating the daughter function and the daughter function with unit energy per symbol in terms of the mother function:
\begin{equation}
\int_{-\infty}^{\infty}\psi^1_{a_i,b_i}(t)\psi_{a_i,b_i}(t)dt = w^1(a)w(a)\int_{-\infty}^{\infty}\psi^2\left(\frac{t-b_i}{a_i}\right)dt
= \frac{1}{\sqrt{Ma_i}}\int_{-\infty}^{\infty}\psi^2(t')dt' =  \frac{\sqrt{M}N_{pr}}{\sqrt{a_i}}.
\end{equation}
\eqnref{eqn:alpha_quadratic_function} is a quadratic function where the quadric coefficient is positive, the function is convex, and thus its minimum is where the derivative is zero:
\[ \frac{dE_{i+1}(a_i,b_i)}{d\alpha_i} =\alpha_i\frac{2MN_{pr}^2}{a_i}-\frac{2\sqrt{M}N_{pr}T_i(a_i,b_i)}{\sqrt{a_i}} = 0. \]
Solving for $\alpha_i$ completes the proof.
\end{IEEEproof}

\bibliographystyle{IEEEtran}
\bibliography{bibtex.bib}

\end{document}